\documentclass[]{aa}  

\usepackage{graphics,graphicx}

\usepackage{hyperref}
\usepackage{natbib}

\usepackage{color}
\usepackage{url}
\usepackage{wrapfig}
\usepackage{amsmath}

\begin{document}

   \title{High-resolution ALMA observations of compact discs in the wide-binary system Sz~65 and Sz~66}

   \author{J. M. Miley\inst{1,2}, J. Carpenter\inst{1}, R. Booth\inst{3}, J. Jennings\inst{4}, T. J. Haworth\inst{5}, M. Vioque \inst{1,6}, S. Andrews\inst{7}, D. Wilner\inst{7}, M. Benisty\inst{8,9}, J. Huang\inst{10},  L. Perez\inst{11}, V. Guzman\inst{12}, L. Ricci\inst{13}, A. Isella\inst{14}
   }
   \authorrunning{J. M. Miley}

   \institute{Joint ALMA Observatory, Alonso de Córdova 3107, Vitacura,
Santiago, Chile.
              \email{jmiley73@gmail.com}
        \and
            National Astronomical Observatory of Japan, NAOJ Chile Observatory, Los Abedules 3085, Oficina 701, Vitacura, Santiago, Chile
        \and 
            School of Physics $\&$ Astronomy, University of Leeds, Sir William Henry Bragg Building, Woodhouse Lane, Leeds LS2 9JT, UK
        \and 
            Department of Astronomy and Astrophysics, Penn State University, 525 Davey Laboratory, University Park, PA 16802
        \and 
            Astronomy Unit, School of Physics and Astronomy, Queen Mary University of London, London E1 4NS, UK
        \and
            National Radio Astronomy Observatory, 520 Edgemont Road, Charlottesville, VA 22903, USA
        \and 
            Center for Astrophysics | Harvard $\& $Smithsonian, Cambridge, MA 02138, US
        \and Universit\'{e} C\^{o}te d'Azur, Observatoire de la C\^{o}te d'Azur, CNRS, Laboratoire Lagrange, F-06304 Nice, France
        \and Universit\'{e} Grenoble Alpes, CNRS, IPAG, 38000 Grenoble, France
        \and 
            Department of Astronomy, Columbia University, 538 W. 120th Street, Pupin Hall, New York, NY, United States of America
        \and
            Departamento de Astronomía, Universidad de Chile, Camino El Observatorio 1515, Las Condes, Santiago, Chile
        \and 
            Instituto de Astrofísica, Pontificia Universidad Católica de Chile, Av. Vicuña Mackenna 4860, 7820436 Macul, Santiago, Chile
        \and
            Department of Physics and Astronomy, California State University Northridge, 18111 Nordhoff Street, Northridge, CA 91330, USA
        \and
            Department of Physics and Astronomy, 6100 Main MS-550, Rice University, Houston, TX 77005, USA
                \\}
    
   \date{Received ; accepted }

 
  \abstract
   {Substructures in disc density are ubiquitous in the bright extended discs that are observed with high resolution. These substructures are intimately linked to the physical mechanisms driving planet formation and disc evolution. Surveys of star-forming regions find that most discs are in fact compact, less luminous, and do not exhibit these same substructures. It remains unclear whether compact discs also have similar substructures or if they are featureless. This suggests that different planet formation and disc evolution mechanisms operate in these discs.}
   {We investigated evidence of substructure within two compact discs around the stars Sz~65 and Sz~66 using high angular resolution observations with ALMA at 1.3~mm. The two stars form a wide-binary system with 6$\farcs$36 separation.
   The continuum observations achieve a synthesised beam size of 0$\farcs$026$\times$0$\farcs$018, equivalent to about 4.0$\times$2.8~au, enabling a search for substructure on these spatial scales and a characterisation of the gas and dust disc sizes with high precision.}
   {We analysed the data in the image plane through an analysis of reconstructed images, as well as in the uv plane by non-parametrically modelling the visibilities and by an analysis of the $^{12}$CO (2--1) emission line. Comparisons were made with high-resolution observations of compact discs and radially extended discs. }
   {We find evidence of substructure in the dust distribution of Sz~65, namely a shallow gap centred at $\approx$ 20~au, with an emission ring exterior to it at the outer edge of the disc. Ninety percent of the measured continuum flux is found within 27~au, and the distance for $^{12}$CO is 142~au. The observations show that Sz~66 is very compact: 90\% of the flux is contained within 16~au, and 90\% of the molecular gas flux lies within 48~au.} 
   {While the overall prevalence and diversity of substructure in compact discs relative to larger discs is yet to be determined, we find evidence that substructures can exist in compact discs. 
   } 


   \maketitle
%

\section{Introduction}

Interferometric imaging of protoplanetary discs at millimetre wavelengths has uncovered a wide and varied range of substructures within the gas, and in particular, within the dust density distribution of protoplanetary discs.  
The most commonly observed substructures are rings and gaps in discs \citep{Andrews2018TheOverview}, as seen in the now famous images of HL Tau \citep{ALMAPartnership2015}, but there are also many instances of azimuthally asymmetric structures, such as spiral arms \citep[e.g. Elias 2-27][]{Perez2016SpiralDisk} or asymmetric dust clumps \citep[e.g. Oph IRS 48,V1247 Orionis][]{vanderMarel2013ADisk, Kraus2017}. 
These substructures are often interpreted to signal the presence of embedded protoplanets or of the mechanisms that lead to planet formation. For example, an embedded planet can carve one or more gaps in a disc, the width and depth of which will depend on the mass of the planet and the properties of the system \citep{Paardekooper2006DustPlanets,Rosotti2016,Dong2017WhatPlanet,Dong2017MultipleSuper-Earth,Nazari2019RevealingObservations}. Although the influence of protoplanets is the most commonly favoured mechanism, other methods have been proposed, including density enhancements at snow lines \citep{Banzatti2015DIRECTBANDS, Zhang2015EVIDENCEDISK}, at the edge of dead zones \citep{Flock2015GapsDisks,Lyra2015RossbyGradients}, or towards concentrations of magnetic flux related to zonal flows in magnetorotation instability turbulence \citep{Bai2014MagneticTurbulence}. Spirals can be produced by embedded planets or by gravitational instability within the disc \citep{Dong2018SpiralInstability}. Clumps or azimuthally extended asymmetries in the dust distribution can form due to dust traps around local pressure maxima, which may have been created by vortices \citep{Ataiee2013AsymmetricEccentricity,Zhu2016, Baruteau2016GasConcentration}. Efficient dust trapping can aid the growth of solid particles by restricting the inward radial drift \citep{Pinilla2012TrappingOuterR}.

The groundbreaking ALMA large program DSHARP resulted in images of 20 protoplanetary discs at high resolution ($\sim$0.035'', or $\sim$5~au), unveiling a staggering array of finely resolved substructure features, including gaps, rings, spirals, and asymmetric clumps \citep{Andrews2018,Huang2018TheSubstructures}. The DSHARP images and other high-resolution observations of protoplanetary discs \citep[see e.g.][]{Dong2018TheDisk,Perez2019LongBaseline} suggest that substructure is certainly very common in class II discs, and that with sufficiently high angular resolution, we should perhaps expect substructure to be ubiquitous.

The DSHARP survey is biased towards brighter discs due to the sample selection methods \citep{Andrews2018TheOverview}, however, and ALMA surveys of gas and dust in star-forming regions have shown that these bright and extended objects are not representative of the entire disc population \citep{Cieza2021TheResolution}. The fainter discs are found to be more radially compact \citep{Tripathi2017ADISKS,Andrews2018ScalingDisks} and are less likely to show resolved substructure \citep[e.g.][]{Long2019CompactCloud}. These so-called compact discs are a significant fraction of the discs that are currently observed, potentially as many as 60$\%$ in the Lupus sample, for example \citep[][]{Miotello2021CompactDisks}. 

Throughout the literature on protoplanetary discs, the term compact disc has come to be used rather loosely to refer to this population of fainter, radially compact discs. This is a very subjective definition because it strongly relies on the angular resolution of the observation. In general, these discs are unresolved, or marginally resolved, by unbiased surveys of star-forming regions of all ages. For an indication of typical size, we can consider the Lupus survey mentioned previously (with an angular resolution equivalent to 30-40~au) \citep{Ansdell2016}, in older discs in Upper Sco (median resolution $\approx$52~au) \citep{Barenfeld2016ALMAAssociation}, and in younger regions such as Ophiuchus, where observations with a resolution equivalent to 28~au detected 133 discs, 60 of which are resolved, and only 8 of which show substructure \citep{Cieza2019TheResolution}.
A growing number of studies focus on these compact discs and their link to detectable disc substructures. From a sample of 32 discs in Taurus with a resolution of $\sim$ 16~au, \citet{Long2019CompactCloud} identified 12 discs with clear structure and 20 smooth discs, 8 of which are associated with wide binaries. 
 
Smooth and structured discs show a similar spread in mass as the stellar host, system age, and disc luminosity. However, a trend has emerged with the radial extent of the disc. In Taurus, the smooth discs have R$\leq$55~au, but the structured discs span a range from 40 - 200~au \citep{Long2019CompactCloud}. 
In a study of very low mass stars that host faint and radially compact discs (90\% of the disc flux is contained within 13-46~au), \citet{Kurtovic2021SizeRegion} reported substructure in three of the six discs in Taurus that they studied with a resolution of 0$\farcs$1, including a wide cavity in the disc of CIDA 1 \citep{Pinilla2021A1}, and hints of unresolved structure in one of the remaining three sources.
On the other hand, high-resolution observations of small dust discs that lack substructure have called into question whether substructure truly is ubiquitous.
For example, based on high-resolution (0$\farcs$055) observations of CX Tau, \citet{Facchini2019HighDisk} reported that the dust disc was relatively unstructured, and that the inner region resembled the inner regions of the DSHARP discs. CX Tau does not possess the large outer rings that are so common in the DSHARP sample, however, and its ratio of gas disc size to dust disc size is high, which could be evidence of an efficient radial drift  \citep{Trapman2019GasEvolution}. Additionally, \citet{Ribas2023TheScales} reported no visible gaps in the dust distribution of the MP Mus protoplanetary disc using a resolution of 4~au.

High-resolution observations are important to identify substructure. Observations with the finest angular resolution will reveal the shortest structures and allow us to place upper limits on the potential size of undetected structure.
For an idea of the scales involved, we can consider that some dust disc substructures have a theoretical maximum size. In order to remain stable, the perturbations in the gas disc must reach at least the gas pressure scale height, H \citep{Dullemond2018}. In protoplanetary discs, H/R typically is 5-10\%. The result is that the dust-trapping rings induced by these gas perturbations occur on spatial scales that increase with R. For inner disc regions, these sizes are comparable to or smaller than the synthesising beam. The gas perturbation itself must be $\approx$ R, but the dust structures that form as a result of these gas perturbations can form even narrower structures \citep{Xu2022TurbulentFormation}. This therefore indicates a maximum size of the substructure, and this size depends on its radial location, which we expect to be resolved angularly for a given observational setup. 
\citet{Bae2022StructuredDisks} discussed this in terms of effective angular resolution, a ratio of $\theta_{\rm disc}$ to the angular size of the disc, and $  \theta_{\rm res}$ is the size of the synthesising beam. They pointed out that the substructure detection rate is 2\% in discs with $\theta_{\rm disc} / \theta_{\rm res}$ $\leq$ 3 and 50\% in discs of 3 < $\theta_{\rm disc} / \theta_{\rm res}$ < 10, but that substructures are found in 95\% of discs with $\theta_{\rm disc} / \theta_{\rm res}$ > 10. For this reason, we must be mindful to ensure that the conclusions we draw on substructure in discs also consider the wider population that has not yet been observed, or has not yet been observed with sufficiently high angular resolution.

An analysis of the interferometric visibilities, rather than the image plane, can aid these investigations. \citet{Jennings2022AAu} hypothesised that the previously considered smooth inner discs of the DSHARP sample may in fact possess more structure than a typical \emph{clean} image reconstruction can detect. Furthermore, a similar analysis of the \citet{Long2019CompactCloud} sample in Taurus revealed new structure within the inner discs \citep{Jennings2022SuperresolutionDiscs}. \citet{Yamaguchi2021ALMAN} used a sparse modelling imaging technique to improve the resolution of the imaging of the T Tau system. In doing so, the authors uncovered a new gap structure in T Tau N and resolved the T Tau Sa/Sb binary.

Understanding the nature of compact discs goes beyond a simple morphological classification of discs because the substructures we identify in protoplanetary discs frequently originate from planet formation in the disc or from disc evolution processes. 
Annular rings and gaps are maintained by dust traps that allow the build up of pebbles at local pressure maxima \citep{Whipple1972OnComets,Pinilla2012RingShape}. In addition to forming annular structures, they prevent rapid radial drift that depletes the outer disc of solid material \citep{Pinilla2012RingShape}. Discs could be drift dominated, in which case solids would be transported inward relatively quickly, or alternatively, they could possess pressure bumps leading to dust traps that prevent the inward drift of solids. 
In this way, pressure bumps may dictate the global evolution of the dust disc by influencing the inward radial drift of solid matter \citep[e.g.][]{Pinilla2012RingShape,Booth2017}, and the inward transport in this manner plays a large role in enriching the inner regions with ices containing frozen-out volatiles, which affects the planet formation in these regions \citep{Schoonenberg2017PlanetesimalOut,Stammler2017a,Booth2019,Banzatti2020HintsDisks}. Establishing to which extent these same processes act in the compact discs through the observable substructures that they manifest enables us to understand in which ways the radially extended and highly structured discs can reflect the wider population. 

Using long-baseline ALMA observations of two class II discs, we test the claim that substructure is ubiquituous in compact discs on spatial scales of $\sim$4~au. The analysis is carried out both in the image plane and in the uv plane of the observed continuum visibilities. We analyse dust continuum emission and molecular gas emission in the context of previous observations towards compact discs and radially extended discs. 

\section{Sz~65 and Sz~66}

\begin{figure*}[ht!]
    \centering
    \includegraphics[width=0.98\linewidth]{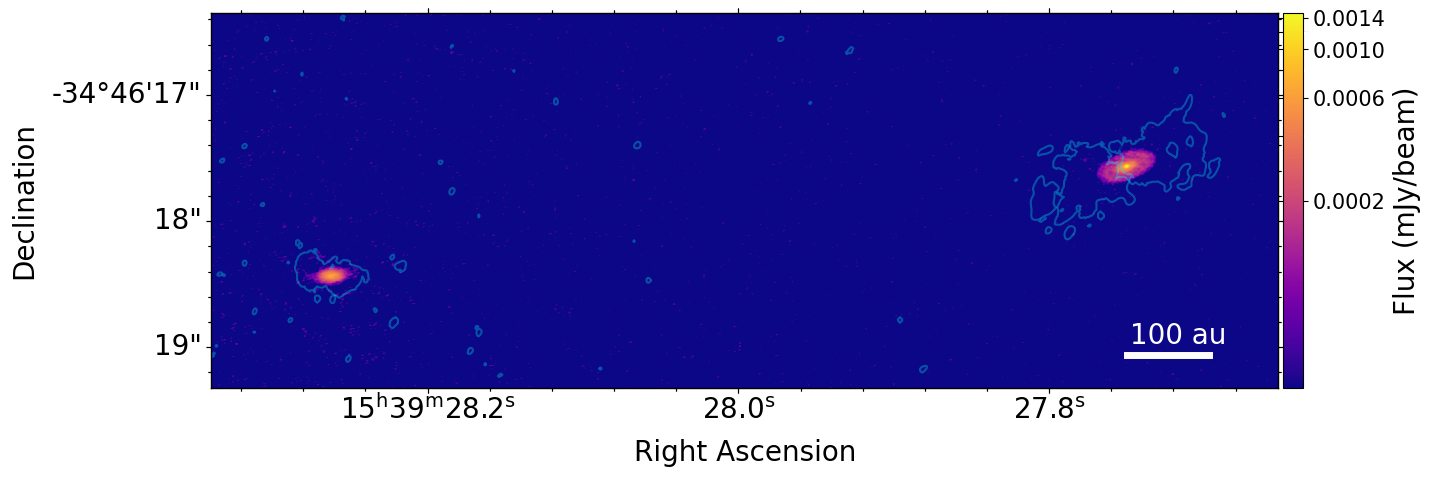}
    \caption{ALMA continuum emission around Sz~65 (right) and Sz~66 (left) (colour map) both shown in the same field. The colour stretch is on a log scale. The contour denotes the gas disc within 5$\sigma_{\rm CO}$. }
    \label{fig:large_field}
\end{figure*}

Our targets were two late-type stars, Sz~65 (spectral type K7) and Sz~66 (spectral type M2), that appear to constitute a wide-binary system because they lie close to each other on the sky. Based on an analysis of X-shooter spectroscopy, stellar masses of 0.76$\pm$0.18~ M$_\odot$ were calculated for Sz~65 and 0.31$\pm$0.04~M$_\odot$ for Sz~66 \citep{Alcala2014X-shooterObjects,Alcala2017X-shooterObjects}. 
Their Gaia distances are very similar, 153.0$\pm$0.6 pc for Sz~65 and 154.4$\pm$0.6 pc for Sz~66 \citep{Bailer-Jones2021Estimating3}, and they are separated by just 6$\farcs$36 on the sky, with Sz~66 97.5$^{\circ}$ East of North from Sz~65 (see also Figure \ref{fig:large_field}). Binary interaction is known to truncate the discs around the stars that are involved, leading to shortened radial extents \citep{Harris2012ASystems,Akeson2014CircumstellarTaurus, Akeson2019ResolvedDisks}, but the past relation of Sz~65 and Sz~66 remains unclear.

The discs were previously observed as part of the Lupus survey with ALMA at 890$\mu$m (angular resolution $\sim$0$\farcs$30), where continuum fluxes of 64.49$\pm$0.32 and 14.78$\pm$0.29 mJy were measured \citep{Ansdell2016}, and at 1.33~mm (angular resolution $\sim$0$\farcs$25), for which the observed integrated continuum flux was 29.94$\pm$0.02 and 6.42$\pm$0.18 mJy for Sz~65 and Sz~66, respectively \citep{Ansdell2016}. \citet{Ansdell2018ALMARadii} measured the dust radius of the well-resolved discs from the Lupus sample, in which Sz~65 is included, but Sz~66 is not. The size of the dust disc in Sz65 was found to be 64$\pm$2~au and the gas radius is 172$\pm$24~au. Sz~65 was modelled in the uv plane by \citet{Hendler2020TheStars}. They reported a smaller dust disc size of 29~au. This size places it within the bottom 38\% of their sample of 199 discs from different nearby star-forming regions, whilst only 7\% of the sample are more compact than Sz~66. \citet{Alcala2017X-shooterObjects} found Sz~65 to be one of the weakest accretors in the sample of 81 Lupus young stellar objects.

\section{Observations}

Our new observations of compact discs around Sz~65 and Sz~66 were taken as part of project 2018.1.00271.S, in ALMA Cycle~6 on 6 July 2019, using 45 antennas across baselines ranging from 149 - 16,196 m, with a mean PWV of 1.2~mm and mean phase fluctuations on the longest baselines of about 150 microns. The observations used three spectral windows optimised for continuum detections with 128 channels across a total window bandwidth of 2 GHz, centred on 232.902, 246.777, and 248.652 GHz, and one spectral window centred on 231.006 GHz, with 3840 channels across a bandwidth of 1.875 GHz, intended for the rotational transition of CO (2-1). The total integration time on source was 88.47 minutes. J1427-4206 was used as the flux and bandpass calibrator (S$_{\nu}$= 1 Jy/beam), J1534-3526 was used as the phase calibrator (1.14$^{\circ}$  from the target source), and J1532-3732 was the check source. All calibrators were also used for WVR corrections.

These new observations were supported by multiple archival ALMA datasets, the details of which we provide in Table \ref{tab:archive}. We concatenate these archival data to our new observations in order to ensure a more complete coverage of the uv plane. This was necessary in particular to fill in the shorter-baseline spacings lacking from the extended configuration of the new observations. These missing baselines that we replaced with archival data trace the more extended emission structures. 

\begin{table*}[h]
\centering
\begin{tabular}{lllll}
\hline
Project Code & Baseline range & PI           & synthesised beam  (")                          & Imaging                        \\ \hline
2018.1.00271 & 149 - 16,196 m & J. Carpenter & 0.024$\times$0.016, -70.8$^{\circ}$ & Dust continuum,  $^{12}$CO (2-1)                  \\ \hline
2015.1.00222 & 15 - 2483 m    & J. Williams       & 0.243$\times$0.233, 49.9$^{\circ}$  & $^{12}$CO (2-1) \\
2017.1.00569 & 15 - 1398 m    & H. Yen       & 0.282$\times$0.204, 73.8$^{\circ}$  & Dust continuum                 \\
2018.1.01458 & 92 - 8547 m    & H. Yen       & 0.046$\times$0.042, -21.0$^{\circ}$  & Dust continuum                 \\
2019.1.01135 & 15 - 784 m     & D. Anderson  & 0.432$\times$0.308, -86.6$^{\circ}$ & Dust continuum                
\end{tabular}
\caption{Details of the ALMA observations. The combined datasets have a representative frequency 230.255~GHz (1.302~mm).}
\label{tab:archive}
\end{table*}

\section{Calibration and data reduction}
The initial data calibration was executed with the ALMA pipeline in CASA version 5.6.1-8. Further data analysis and image reconstruction used releases of CASA version 6. Continuum images were reconstructed from measurement sets with data averaged over all line-free spectral channels in each spectral window. Images were reconstructed using the \emph{clean} algorithm  with Hogbom PSF calculation \citep{Hogbom1974ApertureBaselines} and Briggs weighting \citep{Briggs1995HighSources}. Both sources were imaged from the same field of view. A range of robust values were used during the imaging to investigate the influence on the morphology of the disc; the images presented here use robust = 0.0 . Primary beam correction was applied to all images. Self-calibration was performed individually on the new Cycle 6 and archival data sets before concatenating the data from each of the observations into a single dataset. For observations with the most extended configurations (projects 2018.1.00271 and 2018.1.01458) , the signal-to-noise ratio is low on the longest baselines, and as a result, adequate solutions during phase self-calibration are not found without flagging the longest baselines, which provide the crucial information about the shortest angular scales on the sky. In these cases, the solutions found during phase calibration were applied, but the flags were not (using the option applymode=`calonly' with the CASA task \emph{applycal}). For the data from more compact configurations, self-calibration is more straightforward. Continuum data from project 2019.1.01135 were included after phase self-calibration (stopping at solint=30s) and amplitude self-calibration. Observations from project 2017.1.00569 were included after phase self-calibration (stopping at solint=120s), but amplitude self-calibration did not offer improvements in image quality. Before we combined the individual observations, each data set was aligned to a consistent centre. This was achieved by redefining the phase centre of the measurement set to match that of the centre of a Gaussian fitted to the continuum image from that dataset. In each case, the shifts were smaller than the shift of the full width at half maximum of the synthesising image beam. The coordinate grid was then redefined to a common direction of J2000 15h39m27.750s -34d46m17.692s, adopted from the peak of the dataset with the highest resolution. The final image was reconstructed from the concatenated measurement set consisting of these four individual observations. The first execution block of project 2017.1.00569 was not included as part of the final concatenated data set because anomalous extended emission was identified in the plots of amplitude against uv-distance. Seen in images, these observations were flagged during the quality assurance stages as having high phase and some antenna issues during observations. The remaining executions were sufficient to fill the uv sampling of our resulting concatenated data set without this additional execution.  

The concatenated continuum image had a synthesised beam of 0$\farcs$026 $\times$ 0$\farcs$018, -75.5$^{\circ}$ using Briggs weighting and robust=0.0. This corresponds to a spatial resolution of 3.7 $\times$ 2.5~au as calculated from Gaia early data release 3 parallaxes. The rms measured in the continuum image was 13~$\mu$Jy/beam. 

$^{12}$CO (2-1) emission was not robustly detected by our new observations at the native angular and spectral resolution. Averaging of the spectral windows from 122kHz channels to 768kHz (1~km/s) channels was required in order to extract significant emission from the $^{12}$CO (2-1) line. The data were combined with data from the archival ALMA Band 6 project 2015.1.00222 as presented in \citet{Ansdell2018ALMARadii}. Keplerian masking was applied during cleaning.  A uv taper of 0$\farcs$07 was applied to the data to produce the images presented here, resulting in a synthesised beam of size of 0$\farcs$101$\times$0$\farcs$079, -41.1$^{\circ}$. The rms noise measured from channels in the $^{12}$CO (2-1) data cube is 1.9~mJy/beam km/s. 
In the appendix, Figures \ref{fig:chans_notaper} and \ref{fig:chans_taper} show the effect that the taper has on the emission in the channels. 
We analysed the image cube and moment maps that resulted from the taper-applied imaging.

\section{Results}
\begin{figure}
    \centering
    \includegraphics[width=\linewidth]{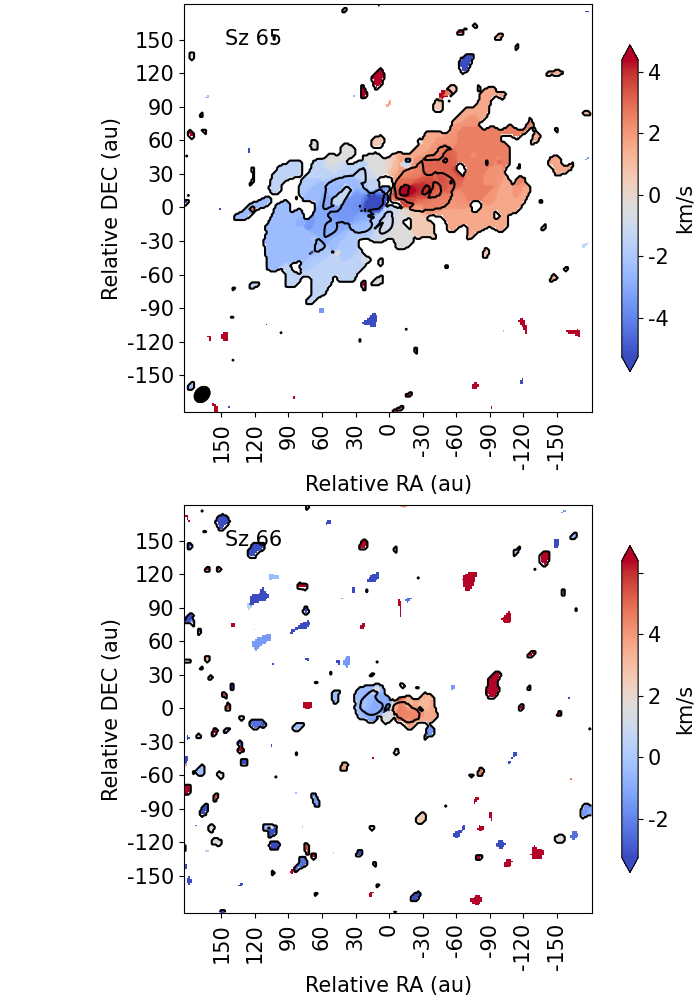}    \caption{Moment 1 map of $^{12}$CO in Sz 65 (top) and Sz 66 (bottom). Contours show the surface brightness in the moment 0 map at (3,15,30) $\times \sigma_{\rm CO}$, where $\sigma_{\rm CO}$=1.3 mJy/beam.km/s. Both maps were created using emission in the cube above a threshold of 3$\sigma$. The Sz~66 field is visibly noisier because it is found towards the edge of the primary beam in the observations that were centred on Sz~65. }
    \label{fig:gas_ims}
\end{figure}

A wide field view of the binary pair is shown in Figure \ref{fig:large_field}. Images with a narrower field of view showing the finer details of the continuum discs are presented in Figure \ref{fig:cont_ims}, and the gas emission is shown in Figure \ref{fig:gas_ims}. 

\subsection{Continuum emission}
\label{sec:Results1_contmaps}

\begin{figure}
    \centering
    \includegraphics[width=\linewidth]{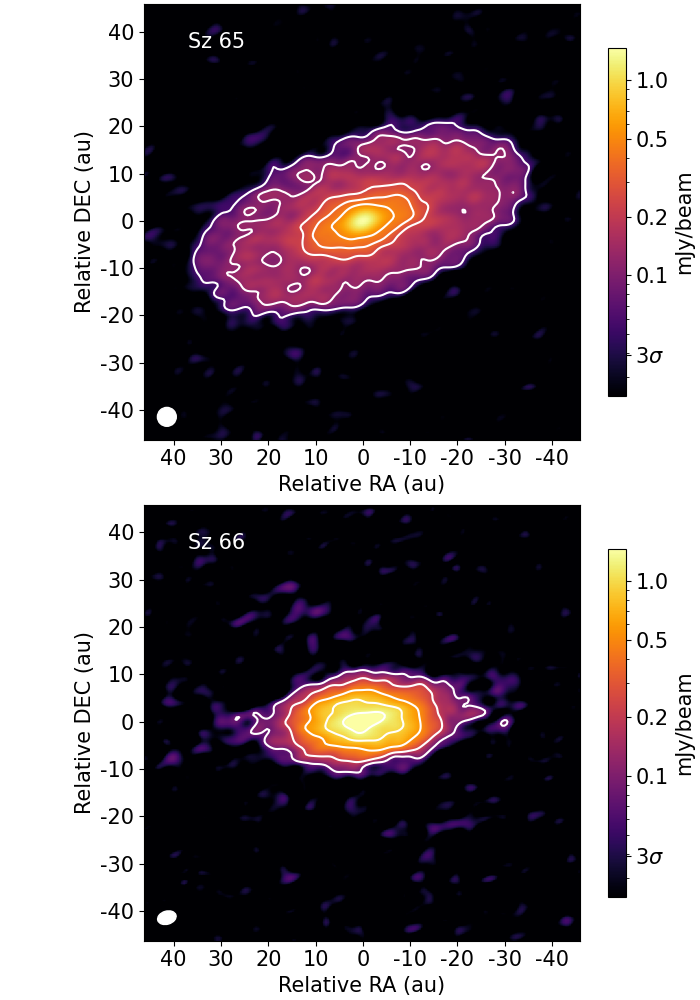}
    \caption{ALMA continuum images of Sz 65 and Sz 66. Contours are shown at (5,10,20,30,40) $\times \sigma_{\rm cont}$. The synthesised beam is shown in the lower left corner by the white ellipse.}
    \label{fig:cont_ims}
\end{figure}

Figure \ref{fig:cont_ims} shows that the emission in the disc of Sz~65 is centrally peaked. Radially exterior to the bright central emission core lies a plateau of fainter emission. A small decrease in surface brightness in this plateau at $\sim$20~au is noticeable in Figure \ref{fig:cont_ims} in between the 10 and 20$\sigma$ contours. We analyse this in greater detail in section \ref{sec:Results2_structure}. The emission does not show any clear azimuthal asymmetries or spiral-like structures. Azimuthally averaging the surface brightness from the disc of Sz~65 gives a 1D intensity profile. This profile reaches the 3$\sigma$ level at 0$\farcs$25 (38~au). When we integrate over this profile out to the point at which the emission drops to the rms of the image, 90\% of the flux from the disc (R$_{90}$) is contained within 0$\farcs$17 (27~au).

The integrated flux density measured within a 3$\sigma$ contour is 28.0~mJy. The peak in emission is 1.34 ~mJy/beam, which for the image rms of 13~$\mu$Jy/beam gives a signal-to-noise ratio of 95 for this observation. A Gaussian fit to the emission in the image plane gives a position angle on the sky of 110$\pm$0.5$^{\rm o}$, a major axis FWHM of 229$\pm$3 milliarcsecond and a minor axis FWHM 103$\pm$1 milliarcsecond, corresponding to an inclination on the sky of 63$\pm$1$^{\rm o}$. Our results are very similar to those found by modelling of the visibilities by \citet{Andrews2018ScalingDisks} using Band 7 observations of Sz~65. \citet{Andrews2018ScalingDisks} reported an inclination of 63$\pm$1$^{\rm o}$, which exactly matches the value we found. They also found a position angle of 107$\pm$1$^{\rm o}$, which is different by 3$^{\rm o}$ from what is derived here. 

For Sz~65, the flux measured in Band 7 (890$\mu$m) by \citet{Ansdell2016} was 64.5~mJy, which gives a spectral index $\alpha=$ 2.3$\pm$0.3, is consistent with the mean spectral index measured in protoplanetary discs \citep{Andrews2020ObservationsStructures}. The uncertainty on the spectral index considers the 10\% flux calibration accuracy achieved for ALMA Bands 6 and 7. 

The dust mass can be estimated from the measured continuum flux by assuming optically thin thermal emission and using Equation \ref{eqn1:dustmass},

\begin{equation}
    F_{\nu} = \frac{\kappa_\nu ~ B_\nu ~ M_{\rm dust}}{d^2},
\label{eqn1:dustmass}
\end{equation}
, where $d$ is the Gaia distance to Sz~65 of 153~pc, $B_\nu$ is the Planck function using frequency of the observation 230.255 GHz and a midplane temperature of 20~K. $\kappa_\nu$ is the opacity of the dust grains, and this is the main source of uncertainty when the dust mass of protoplanetary discs is calculated because it depends upon the grain size, the grain size distribution, and the composition of the dust. We adopted the commonly used prescription $\kappa_\nu = 0.1~ ( \nu / 10^{12}~\rm Hz) ^\beta ~\rm cm^2 g^{-1}$ following \citet{Beckwith1990AObjects}, where $\beta$ is the opacity power index, which we assumed to be 1.0.

With these inputs, we calculate a dust mass estimate of 16.1~M$_\oplus$, which is consistent with the range reported in surveys of Lupus by \citet{Ansdell2016,Ansdell2018ALMARadii}, where dust disc masses of  15~M$_\oplus$ and 20~M$_\oplus$ were derived from Band 7 observations and Band 6 observations, respectively.

We also measured the properties of the continuum emission in Sz~66 from the same image. The integrated flux density is 14.4~mJy, with a peak value of 0.55~mJy/beam, giving a peak signal-to-noise ratio of 39. A Gaussian fit to the image plane returns a best fit with a position angle 80.0$^{\rm o}$ and a shape of 116 mas x 43.8 mas FWHM, corresponding to an inclination of 67.8$^{\rm o}$. These are similar to results for Band 7 observations of Sz~66 with a coarser angular resolution that were analysed by \citep{Andrews2018ScalingDisks}, who found an inclination 69$^{+12}_{-19}$ and a position angle 51$^{+12}_{-19}$. With Equation \ref{eqn1:dustmass}, this gives a dust mass estimate for Sz~66 of 3.4~M$_{\oplus}$. \citet{Ansdell2016} measured a flux at 890~$\mu$m of 14.78~mJy, which led us to calculate $\alpha=2.2 \pm 0.3$. This closely agrees with the value found for Sz~65. R$_{90}$ for the Sz~66 dust continuum is found at 0$\farcs$11 ($\sim$ 16~au). This very compact disc does not show evidence of substructure on scales of 4~au. 

\subsection{$^{12}$CO (2-1) emission}

The rms noise achieved in the line-free channels of the data cube was measured at 1.81 mJy/beam.km/s in the channels of 1km/s width. In Figure \ref{fig:gas_ims} we show the contours of the integrated-intensity moment 0 map of both Sz~65 and Sz~66 overlaid on the moment 1 map. Moment maps are created by combining channels containing line emission above 3$\sigma$. In the moment 0 map, the integrated flux measured from pixels above 3$\sigma$ around Sz~65 is 629~mJy/beam.km/s, with a peak brightness of 61~mJy/beam.km/s. The rms measured from emission-free regions of the moment 0 map created without any noise threshold is 10~mJy/beam.km/s . Around Sz~66, an integrated flux of 382~mJy/beam.km/s is measured and the peak lies at 74.3~mJy/beam.km/s. 

\subsection{Structure of the discs around Sz~65 and Sz~66}
\label{sec:Results2_structure}

\begin{figure}[h]
    \centering
    \includegraphics[width=\linewidth]{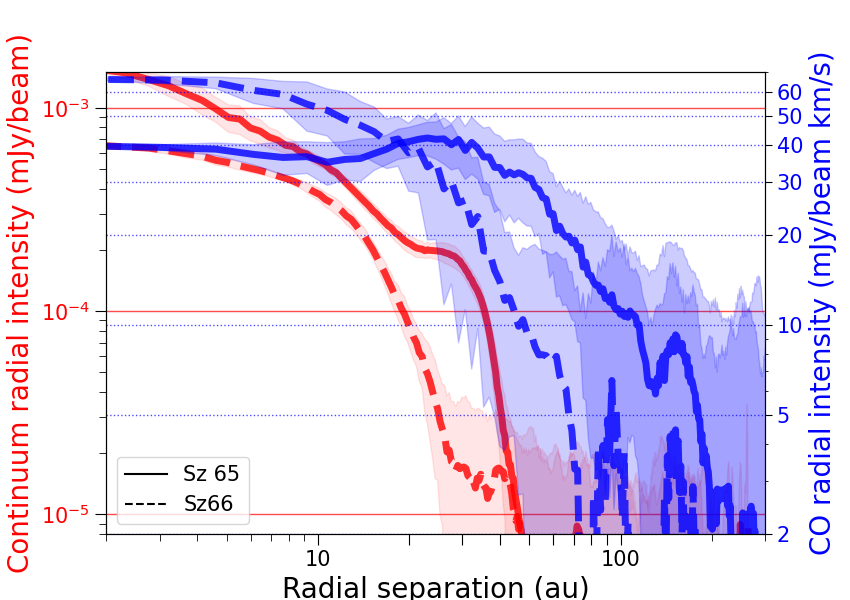}
    \caption{Azimuthally averaged deprojected radial intensity profiles of the dust continuum (red) and CO emission (blue) for Sz~65 (solid line) and Sz~66 (dashed line). The standard deviation in each of the radial bins used for averaging is shown as the shaded area. The beam size for the continuum data corresponds to $\sim$4~au, and the CO data have $\sim$15~au }
    \label{fig:rad_prof}
\end{figure}
The new observations employ longer baselines than previous observations of these discs with ALMA (Table \ref{tab:archive}). This provides new levels of detail in the imaging of Sz~65 and Sz~66 than were previously achieved at these wavelengths. We analysed the measured brightness distribution of dust and gas for evidence of substructure in the compact discs. 

\subsubsection{Measurements from the image plane} 
\label{sec:meas_from_im}

In Figure \ref{fig:cont_ims} we resolve the disc of Sz~65 well enough to examine the radial structure. The FWHM of a Gaussian fitted to the continuum image is $\sim$9.5 times the FWHM of the synthesised beam. In the case of Sz~66, about five beams fit across the major axis. The image of Sz~65 shows a bright central core in the centre of an extended disc of weaker emission. This is seen more clearly when the emission is deprojected using the inclination and position angle of the disc that we defined from a Gaussian fitted to the continuum emission. We then azimuthally averaged the emission to produce radial flux profiles, which are shown in Figure \ref{fig:rad_prof}. The properties of the two discs extracted in this section are summarised in Table \ref{tab:im_results}.

In Figure \ref{fig:rad_prof} we compare the extent of the dust disc to that of the molecular gas disc using CO(2--1) as a tracer. In order to compare the extension quantitatively, we also defined the disc size, R$_{90}$, as the radius within which 90\% of the cumulative flux measured from the image plane is contained. The uncertainties correspond to the width of the bins of the deprojected radial distance, which is related to the pixel size of the reconstructed images. R$_{90}$ for the continuum is 27$\pm$4~au, and for the CO gas, it is 142$\pm$15~au around Sz~65. We find 16$\pm$4~au and 47$\pm$15~au in Sz~66. The uncertainties in disc sizes are given as the deprojected beam size in au. The decrease in intensity of CO towards the inner regions of Sz~65 is thought to be due to the effects of continuum subtraction. 
Compared with the results of \citet{Ansdell2018ALMARadii}, who achieved an angular resolution of $\sim$36~au with archival data, our R$_{90}$ for the CO emission is consistent within the uncertainties with their value of 172$\pm$24~au. \citet{Ansdell2018ALMARadii} found a larger size of 64$\pm$2~au for previous continuum observations with a coarser angular resolution, but our value is consistent with measurements made in analyses in the uv plane by \citet{Andrews2018ScalingDisks} and \citet{Hendler2020TheStars}.

The ratio of the observed gas and dust disc sizes can be indicative of the physics that drove the previous disc evolution. Thermochemical modelling by \citet{Trapman2019GasEvolution} posited that R$_{\rm gas}$/ R$_{\rm dust} >4$ is a clear signal that there has been significant dust evolution within protoplanetary discs. 
In the sample of \citet{Long2022GasEvolution}, who investigated gas disc sizes, 8 of 44 discs have large R$_{\rm gas}$/ R$_{\rm dust}$, that is, $>4$, suggesting strong radial drift. 
Of the six very low mass stars with compact discs studied by \citet{Kurtovic2021SizeRegion}, four had a ratio greater than or close to four. The authors noted, however, that these measurements were conservative constraints because cloud contamination in Taurus can be strong. 

The ratio of R$_{\rm gas}/$R$_{\rm dust}$ for Sz~65 is 5.3$\pm$0.9. This high ratio may indicate significant grain growth and radial drift in the disc \citep{Trapman2020ConstrainingLupus}. A high R$_{\rm gas}/$R$_{\rm dust}$ can also be explained by optical depth effects \citep{Facchini2017}, but we resolved the disc very well and the ratio of 5.3 is particularly high, suggesting that this is less likely to be the case for Sz~65. For Sz~66, the ratio is lower at 2.9$\pm$1.0, consistent with that of single-star discs in Lupus and Taurus \citep{Ansdell2018ALMARadii, Long2022GasEvolution}. 
We know that the disc is part of a binary system, even in a system with a relatively large separation. The truncation of the disc during a previous close encounter between the two binary partners might therefore be the cause. The relative uncertainties in the size estimation are high because the disc is spatially less well resolved in the case of the gas. Observations with a higher signal-to-noise ratio in the outer regions of the disc would be able to place a more precise constraint on these results.

The gas and dust profiles in Figure \ref{fig:rad_prof} do not show multiple successive rings and gaps, but are instead fairly smooth. These intensity profiles were deprojected using the inclination and position angle given in Section \ref{sec:Results1_contmaps}. The dust continuum emission from Sz 65 on a logarithmic scale (Figure \ref{fig:rad_prof}) shows one bump followed by a plateau at the outer edge of the disc $\sim$20~au, but there are no deep gaps as seen in some DSHARP images, for example. This behaviour is similar to that of other compact discs that were observed at high resolution with ALMA, as we explore in Section \ref{sec:DSHARP_comp}. The radial profile of the CO emission in Sz~66 is smooth until it reaches the noise levels. The central part of the gas disc is very bright, suggesting unresolved and optically thick emission at short separation from Sz~66. We note again here that a uv taper of 0$\farcs$07 was applied to the data. Even deeper observations in the future may be able to reveal the details of these inner regions with more clarity. 

\begin{table}[]
\centering\begin{tabular}{l|ll}
\textbf{Disc properties}                 & \textbf{Sz 65}       & \textbf{Sz 66 }     \\ \hline
mm flux (mJy)            & 28$\pm$3    & 6$\pm$1   \\ 
R$_{90}$ dust (au)       & 27$\pm$4    & 16$\pm$4   \\
R$_{90}$ gas (au)        & 142$\pm$15  & 48$\pm$15  \\
Inclination (degrees)    & 63$\pm$1    & 68$\pm$1   \\
Position angle (degrees) & 110.0$\pm$1 & 80.0$\pm$1
\end{tabular}
\label{tab:im_results}
\caption{Basic disc properties for Sz~65 and Sz~66.}
\end{table}

\subsubsection{Non-parametric visibility fitting}
\label{sec:FRANK}

To scrutinise the radial brightness structures in Sz~65 further, we made a fit to the visibility data using the super-resolution fitting code \textsc{frank} \citep{Jennings2020FRANKENSTEIN:Process}\footnote{https://discsim.github.io/frank/}. 
We used the log-normal fitting method in \textsc{frank,} which fits for the radial profile in logarithmic brightness space. This version applies the power spectrum prior to log(I$_\nu$), and the fitting procedure follows that described in \citet[][i.e. RESOLVE]{Junklewitz2016RESOLVE:Astronomy}. The log-normal version results in a smoother brightness profile with radius, as the fit is less prone to fitting rings of small amplitude or in noisier baseline ranges. 

\begin{figure}[h]
    \centering
    \includegraphics[width=\linewidth]{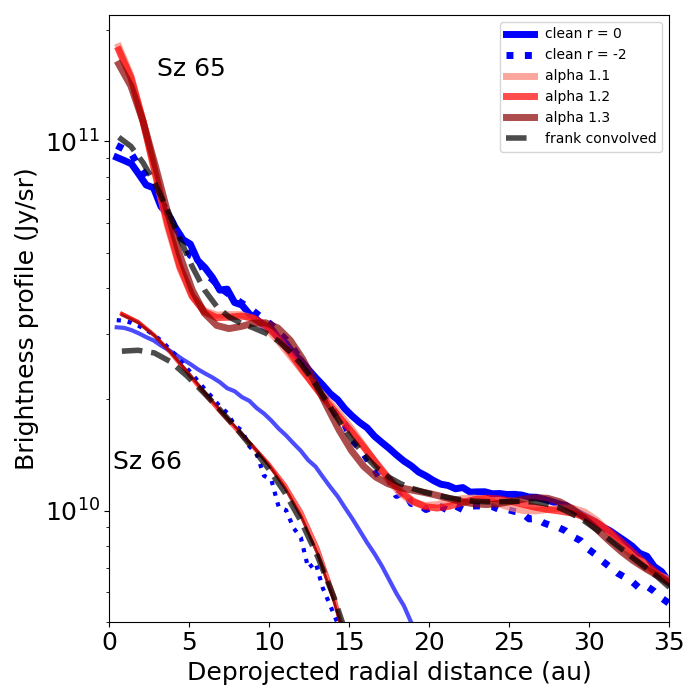}
    \caption{Azimuthally averaged brightness profile from reconstructed images of Sz~65 and Sz~66 using robust=0 and robust=2. We also plot the brightness profile found by the non-parametric visibility fitting. We show three different model fits with a varying fit parameter $\alpha$. Additionally, we show the \textsc{frank} profile convolved with the synthesised beam of the observations, and a profile derived from an image using robust=-2. }
    \label{fig:clean_v_frank}
\end{figure}

\begin{figure}
    \centering
    \includegraphics[width=\linewidth]{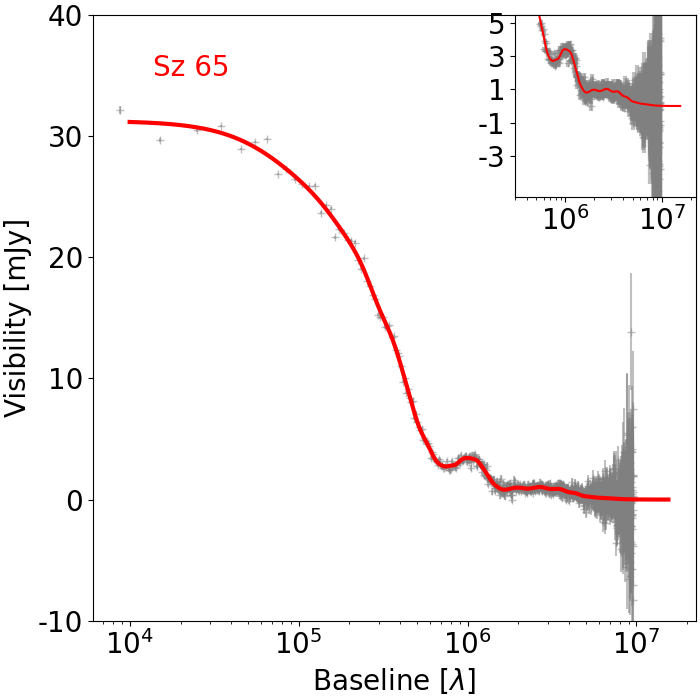}
    \caption{Model fit to the real component of the deprojected visibilities of Sz~65 as a function of the baseline. }
    \label{fig:clean_v_frank_uv}
\end{figure}


When multiple emission sources lie in the field of view of an observation, each contributes to the measured signal. Sz~66 is 6$\farcs$36 from Sz~65 and contained within the field of view of our observations, and so we must proceed cautiously with the analysis in the uv plane. 
Fortunately, Sz 65 is significantly brighter than Sz 66, and the angular separation between the two corresponds to a drop in efficienty to 0.84 in the primary beam of our observations. We verified that our non-parametric fit to the visibility plane successfully isolated only Sz 65 by fitting a model to Sz 66, and subtracted this from the visibilities. 
The brightness profile we extracted from the dataset from which the emission of Sz~66 was subtracted is indistinguishable from the profile extracted from the original data. The fainter emission at a large separation from the phase centre, such as that belonging to Sz~66, therefore has a negligible effect on the model fit. The features we see are therefore expected to correspond to the continuum emission of Sz~65 alone. 

Figure \ref{fig:clean_v_frank} compares the average 1D brightness profile obtained from \emph{clean} images with non-parametric fits to the visibilities using \textsc{frank}. Two \emph{clean} profiles are shown, one using the fiducial robust=0.0 as shown in Figure \ref{fig:cont_ims}, and one using robust=-2.0. Robust = -2.0 gives more weight to longer baselines, which are sensitive to the smallest spatial scales. This increases the resolution at the expense of sensitivity. This profile is very similar to that achieved using robust =0.0, except in outer regions, where noise becomes more significant. Here, the robust=-2.0 image results in a shorter disc extension. For each disc, we used the disc geometry parameters given in Table \ref{tab:im_results} to deproject the emission. The maximum radius was set to 2$\farcs$, and the number of collocation points used in the fit, N, was set to 300. In Figure \ref{fig:clean_v_frank_uv} we show the fit in visibility space, where the model fit line follows the data points very closely.
When $\alpha$ is varied, the \textsc{frank} hyperparameter related to the signal-to-noise ratio threshold for visibilities to be included in the fit does not affect the result. In Figure \ref{fig:clean_v_frank} we show fits using various alpha values to demonstrate this. For the remainder of this work, we refer to the alpha = 1.2 case as the `\textsc{frank} fit'. We also varied the wsmooth \textsc{frank} parameter that acts to smooth the profile that is obtained from the fitting, but these variations did not significantly affect the form of the radial profile. 
A slight adjustment to the position of the disc was made relative to the Gaussian fit achieved as part of the \textsc{frank} fitting. The \textsc{frank} modelling begins with a Gaussian fit to the emission made in the uv plane in order to determine the geometry of the system. This determines the offset in RA and DEC from the phase center, the position angle, and the inclination of the disc. This method finds an inclination = 63.8$^\circ$, a position angle = 110.2$^\circ$, dRA= -0.54 mas, and dDec= -2.23 mas. The inclination and position angle determined as part of the visibility analysis agree with fits to the image plane.
We ran a wider grid of models with the central position of the disc shifted by up to $\pm$5 mas to check for reduced residuals, as demonstrated in \citet{Andrews2021LimitsDisks,Jennings2022SuperresolutionDiscs}, for example. The case with the least residuals is found to be a fit with a shift in space (dRA$+$1mas, dDec$-$4mas).

Overall, the model and the image identify similar structures in the dust disc of Sz~65, as shown by Figure \ref{fig:clean_v_frank}. They retrieve the same plateau at about 20~au, which continues until the edge of the disc before a sharp drop. The visibility fitting suggests a smaller feature at about 6~au in Sz~65, for which there is no counterpart in either of the profiles extracted from the image plane. We therefore remain cautious about this potential gap. The visibilities tracing these scales are those measured on the longest baselines, for which the signal-to-noise ratio is lowest. In this case, future observations of the system are required to confirm the veracity of this feature, whether through higher signal-to-noise ratio millimeter imaging or at shorter wavelengths, where a higher angular resolution can be achieved.
We can parametrise the gap size following \citep{Long2018GapsRegion,lodato2019} as the separation between the brightness minimum in the gap and brightness peak in the ring. The local brightness minimum in the visibility fitting model is at a deprojected separation of 20~au, and the peak brightness of the exterior ring is at 24~au. This gives a gap a size of 4~au, or a full width of 8~au assuming azimuthal symmetry. The brightness difference between these two locations is 8\%.
Convolution of the fitted model with a Gaussian of the same size as our synthesised beam reduces the amplitude of the inner regions slightly and smooths out this potential structure. With this exception, the image plane at very short separation from the star gives a very similar brightness distribution as the non-parametric fit to the visibilties. This can be described as a bright inner disc with a gap-ring structure in the outer disc.

It is possible that the general consistency between the image plane and uv plane profiles in Figure \ref{fig:clean_v_frank} arises because Sz~65 does not host any finer (unseen) structure in its disc, and the angular resolution of our observations might be sufficient to trace the density variations in the disc, or because any structures that do exist are on scales much smaller than $\approx$~3~au. We measured a continuum R$_{90}$=27~au, and therefore, this should perhaps be expected as the scale height, H, at the very edge of the dust disc is likely $\leq$ 2.7~au. In this case, stable structures caused by pressure bumps in the gas disc that in turn shape the dust distribution should be smaller than our angular resolution in the inner regions of the disc.  

\begin{figure}
    \centering
    \includegraphics[width=\linewidth]{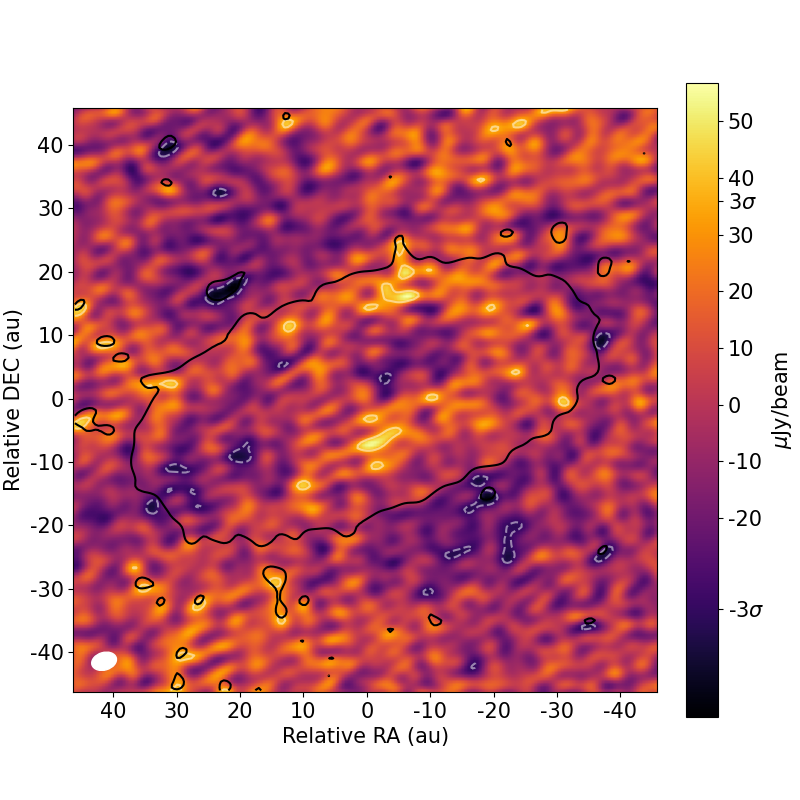}
    \caption{Residual emission for Sz~65 following the subtraction of an axisymmetric model fitted to the visibilities from the original data. The solid white lines show residuals in steps of $\sigma$, positive contours begin at 3$\sigma$, and dashed lines show negative contours that begin at -3$\sigma$. The solid black line shows the 3$\sigma$ emission contour of the continuum image prior to the model subtraction.}
    \label{fig:frank_resid}
\end{figure}

In Figure \ref{fig:frank_resid} we show the residual emission that remains after the \textsc{frank} model is subtracted from the \emph{clean} image of Sz~65. 
Only some very low level residuals remain, which barely exceed the 3$\sigma$ level measured from this map. This residual map shows that the \textsc{frank} fit is a very good representation of the data down to the noise level. 


\begin{figure}
    \centering
    \includegraphics[width=\linewidth]{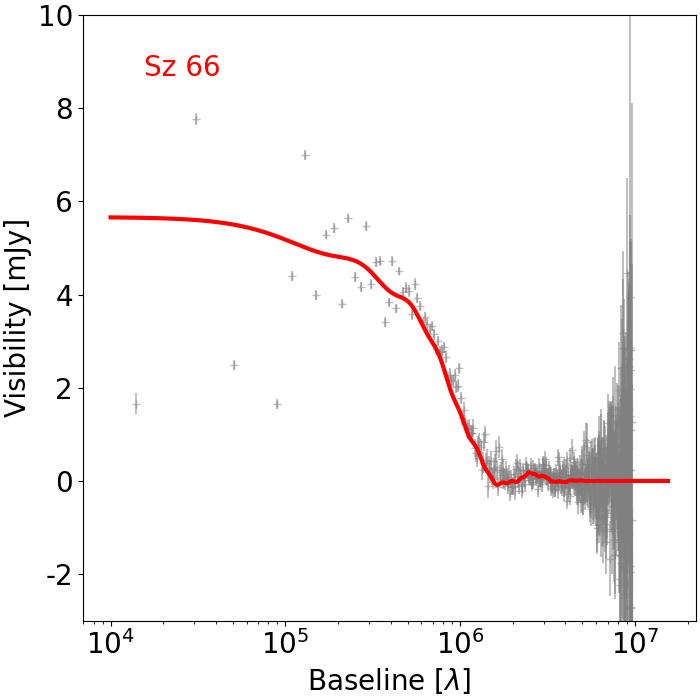}
    \caption{Model fit to the real component of the deprojected visibilities of Sz~66 as a function of the baseline. }
    \label{fig:clean_v_frank_uv_66}
\end{figure}

\begin{figure}
    \centering
    \includegraphics[width=\linewidth]{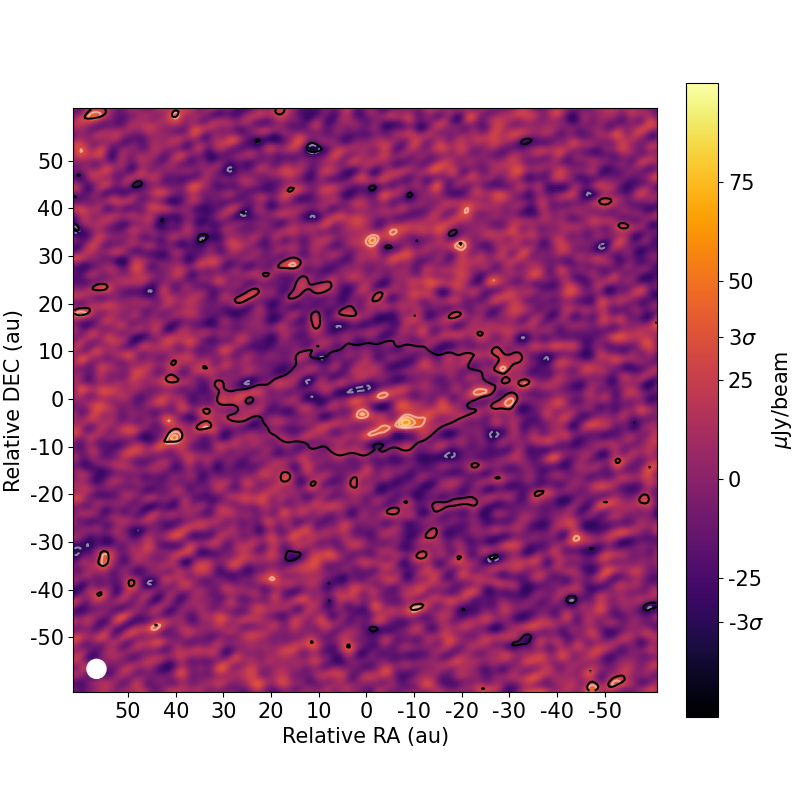}
    \caption{Same as Figure \ref{fig:frank_resid}, but for Sz~66.}
    \label{fig:frank_resid_Sz66}
\end{figure}

We repeated this process for the disc of Sz~66, for which the comparison of the model with the visibilities is shown in Figure \ref{fig:clean_v_frank_uv_66}. The residual map is shown in Figure \ref{fig:frank_resid_Sz66}. The \textsc{frank} model describes the emission well generally, although localised residual emission that reaches 7$\sigma$ can be seen towards the south-west of the image, perhaps indicating an asymmetry in the outer regions of the disc.

\subsubsection{Analysis of the molecular gas disc}
\label{sec:gas}

We analysed the molecular gas disc of Sz~65 using our imaging of CO(2-1) as shown in Figure \ref{fig:gas_ims}. We note that the CO brightness line decreases around the line centre, primarily along the minor axis of the discs, possibly due to an oversubtraction of continuum. This limits our ability to search for misalignments between the inner and outer disc. Misalignments can be driven by massive embedded planets on inclined orbits \citep{Nealon2018WarpingOrbit}, via fly-bys, or through interactions with a companion \citep{Cuello2019FlybysDynamics}. If the discs of Sz 65 and Sz 66 have interacted in the past, we may expect to see some of the more common observational signatures such as spirals, disc truncation, or disc warping in observations of the CO line \citep{Cuello2020FlybysSignatures}. If the disc was warped by the binary interaction, we would expect to see a departure from Keplerian rotation in the velocity map of the disc. The subsequent velocity perturbations are generally about 0.1--1 km/s \citep[see e.g.][]{Wolfer2023KinematicsALMA}, dependent upon the properties of the flyby of the binary partner, such as the inclination relative to the disc rotation, the separation distance, and the mass ratio of the partners \citep{Nealon2018WarpingOrbit,Cuello2020FlybysSignatures}. Our line cubes were averaged to a width of 1km/s. A finer spectral resolution is typically required to detect deviations from a Keplerian disc, unless the disruption has been particularly dramatic. 

The continuum models from the visibility fitting were derived from a 1D radial profile, meaning that they assumed axisymmetry, but the spirals or perturbations induced by the influence of an external companion are not expected to be axisymmetric \citep[e.g.][]{Kurtovic2018TheSystems}. They can be studied by using gas line observations, however, to search for perturbations to the velocity structure. 

\begin{figure}
    \centering
    \includegraphics[width=\linewidth]{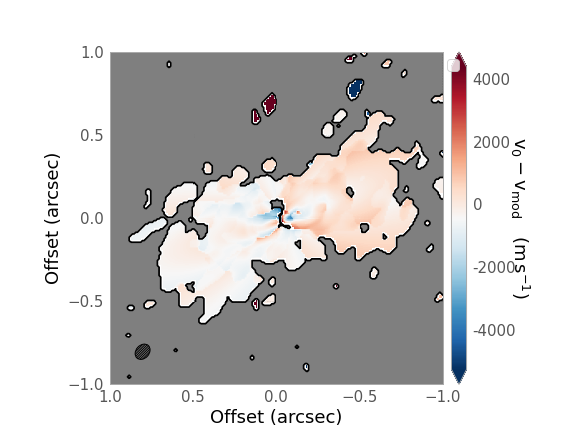}
    \caption{Residual map following the subtraction of a Keplerian disc model with elevated emission constructed from the velocity information in $^{12}$CO observations towards Sz~65. The colour scale is set to match the full range of the data before subtraction.}
    \label{fig:Sz65_eddy_resid}
\end{figure}

We used the tool \emph{eddy} \citep{Teague2019EddyDYnamicse} to fit a rotation model to each of Sz~65 and Sz~66 using a model of a gas disc that rotated in a Keplerian manner. We subtracted this model from the map of peak velocities of the cube created with \textit{bettermoments} \citep{Teague2018ALines} using an image created with Briggs weighting and robust=0.0, a spectral resolution averaged to 1.0~km/s, and a 3$\sigma$ threshold. The residual map following the subtraction of the model is shown in Figure \ref{fig:Sz65_eddy_resid}, where the colour stretch is the same as in Figure \ref{fig:gas_ims}. The peak residuals following the subtraction of the model from the Sz~65 moment 1 map reach $\sim$ 1.3 km/s within the main disc. We assumed that the emission outside of the central 3$\sigma$ emission is noise. These deviations are relatively small considering the initial spectral resolution of the cube 1.0 km/s. We are only sensitive to quite significant disruptions to Keplerian motion, such as from interaction with a binary partner. The residuals with the greatest magnitude are confined to the inner regions. 
Through this modelling, we also obtained a kinematic constraint on the stellar mass of Sz~65 of M$_* =0.77 \pm 0.01$~M$_\odot$. This agrees very well with the results of spectroscopic modelling by \citet{Alcala2017X-shooterObjects}, who derived a mass of 0.76~M$_\odot$. For the final fit to the disc, we fixed the robust mass and inclination values of the disc to avoid issues that can arise due to the degeneracy between the parameters (see Figure \ref{fig:corner65} for the results of the MCMC fitting). The kinematic mass of the central object can be constrained with greater accuracy by using observations of greater angular and kinematic resolution. The disc self-gravity and stellar contribution to the gravity potential might also be taken into account where appropriate, as in \citet{Veronesi2021A227}. We included vertical parameters in the fitting in order to constrain the vertical height of the disc. We used \emph{eddy} to assume an azimuthually symmetric emission surface described by 
\begin{equation}
    z(r) = z_0 \times \Bigg(\frac{r}{1"}\Bigg) \times {\rm exp} \Bigg(\frac{r}{r_{\rm taper}}\Bigg)^{q_{\rm taper}}
\label{eqn:elev}
\end{equation},

where r is the radial separation from the star ,and $\psi$ describes the disc surface flaring. r$_{\rm taper}$ and q$_{\rm taper}$ describe the exponential taper at the edge of the disc where the emission height drops off. This modelling achieves a best-fit value of the aspect ratio constant z0=0.03 (see Figure \ref{fig:corner65} for the MCMC results). This result is undoubtedly affected by the uv taper and spectral averaging applied to the data, and so this figure should represent a firm lower limit. Future observations achieving higher resolution both spatially and spectrally may reveal a higher surface layer in the disc. Despite introducing the elevated emission surface, the residuals on the right-hand side of the plot appear to be slightly redshifted, suggesting an undersubtraction in this area. A grid of models in which PA and inclination were varied in the fitting process did not reduce this effect, suggesting that it is not a issue that arises due to a poor fit to the disc geometry. It is most likely that these slightly redshifted residuals are a result of the relatively coarse spectral resolution of the observation of 1km/s. Alternatively, there are physical explanations that might contribute to this effect. The undersubtraction by the model might indicate an inability to achieve a close fit to the vertical emission height as parametrised in Equation \label{eqn:elev}, or it could be the result of an overdensity in the disc, as has been modelled in circumbinary systems by \citet{Calcino2023ObserN}.

\begin{figure}
    \centering
    \includegraphics[width=\linewidth]{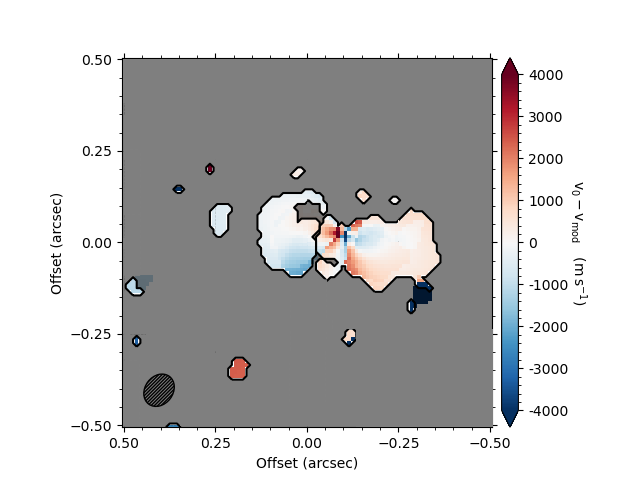}
    \caption{Same as in Figure \ref{fig:Sz65_eddy_resid}, but for Sz~66. The $^{12}$CO detection in this disc is much weaker and affects the model fit.}
    \label{fig:Sz66_eddy_resid}
\end{figure}

In Figure \ref{fig:Sz66_eddy_resid} we show a similar modelling of Sz~66, this time without invoking an elevated emission surface (the MCMC results are shown in Figure \ref{fig:corner66}). This fitting procedure finds an inclination value for the $^{12}$CO gas disc of 38$^\circ$, which is significantly different from that derived from images of the dust disc, which was found to be 68$^\circ$ . We note, however, that it is more difficult to achieve a good Keplerian fit to the data for Sz~66. The residuals are greater in Figure \ref{fig:Sz66_eddy_resid} because the data are noisier and because the disc is less well resolved spatially. Large residuals on both the blue- and redshifted side of the disc can be seen that perhaps suggest a vertical component to the gas disc, but the angular resolution of the observations is not sufficient to resolve this difference given the size of the disc we measure. It is difficult to discern true physical effects due to the low signal of the detection in Sz~66, and so we refrain from placing any further constraints on the structure from a kinematic analysis.

The dust emission (e.g. Figure \ref{fig:large_field}) contains no evidence of entrained dust material between the two discs that might suggest significant binary interaction or a recent fly-by. No detection is made, even though we included observations made with the more compact configurations, which are sensitive to such large spatial scales (Table \ref{tab:archive}). The maximum recoverable scale of the most compact configuration included in these observations, project code 2019.1.01135, is 5$\farcs$8.
There does exist VLT/NACO imaging of the binary \citep{Zurlo2021TheRegion}, in which there is bright emission around Sz~66 (tracing dust coupled to the gas) that might support extended diffuse emission, but no arc or spiral arm structure is apparent.

\subsubsection{Tidal truncation}

We can use the constraints on disc size to investigate the likelihood of a previous tidal truncation of the protoplanetary discs through binary interaction. Theoretical models estimate the truncated equilibrium tidal radii, R$_t$, for discs in a binary pair that is dependent upon the orbital parameters, such as the separation between the two sources, $a$, the eccentricity of their orbit, $e$, and the mass ratio of the two stars, $q$. The measured size of a protoplanetary disc can then be compared with the predicted tidal truncation radius for wide or intermediate separation binary pairs, as demonstrated in \citet{Harris2012ASystems,Panic2020PlanetSystems}. 
\citet{Pichardo2005CircumstellarBinaries} gave a semi-analytical approximation for a truncated disc size that we can write as
\begin{equation}
    R_t \approx 0.337 \Bigg[ \frac{(1-e)^{1.2} q ^{2/3} \mu ^{0.07}}{0.6 q^{2/3} + ln(1+q^{1/3})} \Bigg] a,
\end{equation}
where $\mu$ is the mass fraction of the stars (= q/(1+q) ).
We do not know the true separation, $a$, of the two stars, but we can instead use the projected angular separation derived from Gaia parallaxes to create a probabilistic model of R$_t$ by following \citet{Torres1999SubstellarApproach}, who gave the ratio of the semimajor axis to the physical separation as 
\begin{equation}
    F \equiv \frac{a}{a_p} = \frac{1-e^2}{a+e~cos(\nu)} \sqrt{1-sin^2(\omega + v)~sin^2i } ,
\label{eqn2:trunc}
\end{equation}
where $\omega$ is the longitude of periastron, and $v$ is the true anomaly. Following the approach of \citet{Harris2012ASystems}, we used a Monte Carlo approach to sample $\omega, i, e $, and $v$, assuming the eccentricity distribution to be uniform, and that orbital inclinations have a $sin(i)$ dependence. This analysis also assumes that the disc midplane lies on the orbital plane of the binary system. The two discs have a comparable inclination on the sky as derived from the continuum observations ($\approx 61^\circ and  68^\circ$, for Sz~65 and Sz~66, respectively). Wide binaries are often found to be slightly misaligned with each other (\citep{Jensen2004TestingBinaries}). If a similar inclination in the individual discs traces an initial inclination of the binary orbital plane, it might support a scenario in which the two systems formed from the same molecular cloud. There are many processes, however, that can lead to a misalignment of the two discs during their formation. The relative alignment in class II may be due to mechanisms acting after formation to re-align the discs \citep{Bate2000PredictingSystems,Jensen2004TestingBinaries}.
Using Equation \ref{eqn2:trunc}, we calculate a probability distribution, P(R$_t$), that each disc has been truncated tidally to a radius R$_t$. We can calculate P(R$_t$) for Sz~65 by defining q= M$_{\rm Sz65}$/M$_{\rm Sz66}$, or alternatively, for Sz~66 by inverting this ratio.  

\begin{figure}
    \centering
    \includegraphics[width=\linewidth]{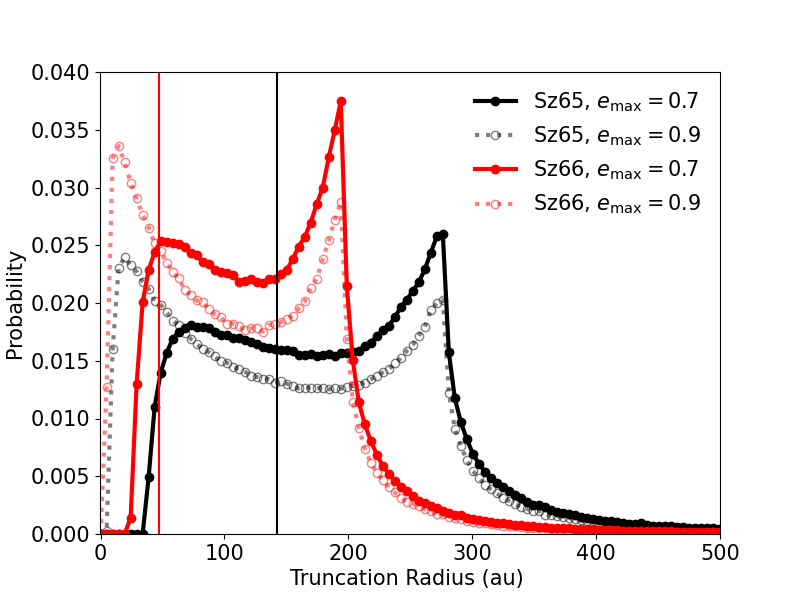}
    \caption{Probability distribution P(R$_t$) for the disc radii expected in Sz~65 and Sz~66 assuming tidal truncation of the disc by its binary partner. The calculations were run for each binary partner for two different uniform eccentricity priors, one with e$_{\rm max}$ = 0.7, and one allowing for high-inclination solutions with e$_{\rm max}$ =0.9. The vertical lines show the measured R$_{90}$ for each disc by this work. }
    \label{fig:trunc_fig}
\end{figure}

Figure \ref{fig:trunc_fig} shows the results of the calculations for a truncation radius of both Sz~65 and Sz~66 after running 10$^6$ models for each case. These samplings were repeated for two different uniform eccentricity priors [0.0, 0.7] as adopted in \citet{Harris2012ASystems} and [0.0, 0.9] in order to consider high-eccentricity cases. By considering high eccentricity in the binary orbit, lower R$_t$ results become more likely. This is shown in Figure \ref{fig:trunc_fig} by the dotted line, which represents a set of solutions calculated by sampling a wider range of eccentricity values with e$_{\rm max}$ up to 0.9. As discussed in  \citet{Harris2012ASystems}, we currently lack the empirical information to characterise the eccentricity distribution among multiple systems with statistical significance. Systems with short orbital periods tend to have low eccentricity as a result of the tidal circularisation of the orbit \citep{Zahn1977TidalStars,Zahn1989TidalBinaries,Melo2001OnPeriod}, but for systems with longer periods, the distribution can be considered relatively uniform for 0.1 < e < 0.9 \citep{Mathieu1994Pre-main-sequenceStars}. 

In Figure \ref{fig:trunc_fig}, Sz~66 is expected to be the more compact disc, with its most likely truncation radius found at 195~au. This is large in comparison to the R$_{90}$=47~au we derive from the $^{12}$CO imaging. Only 7\% of the solutions for R$_t$ are smaller than the derived R$_{90}$ for Sz~66. As shown in Section \ref{sec:meas_from_im}, Sz~66 has a gas/dust size ratio similar to that of single stars in surveys of star-forming regions, and overall, it therefore seems unlikely that tidal truncation causes such a compact disc. The disc may have formed radially compact from the very beginning.
Sz~65 has a most likely predicted truncation radius of 276~au, which is also much larger than the measured R$_{90}$, which is 161~au. In Figure \ref{fig:trunc_fig} the vertical line indicating our derived R$_{90}$ is towards the middle of the range of predicted values. From this analysis, there is not sufficient evidence that tidal truncation is the primary influence in setting the disc sizes we measure.

\section{Discussion}
\label{sec:Discussion}
\subsection{Evidence of substructure in the high-resolution observations of Sz~65 and Sz~66}

One of the main results of the DSHARP survey suggested that substructure in large protoplanetary discs was ubiquitous. We have presented high-resolution observations of two compact previously poorly resolved discs in order to search for substructure. The two discs are compact and have low-mass discs surrounding late-type stars in a wide binary system. Both discs are well resolved by the achieved synthesised beam, which allows a search for substructures in the dust discs. Highly substructured discs would suggest dust traps and would prevent the continued radial inward drift of the dust grains. If they are structureless on smaller spatial scales as well, this might suggest that radial drift works efficiently and that there is a lack of dust trapping at large radii. 
To confirm the trapping scenario for certain would require resolving the trapping scale of $\sim$H. For these systems, this confirmation remains tantalisingly out of reach; at the measured R$_{90}$ for Sz~65, a typical scale height in protoplanetary discs of H $\approx$ 0.1R would give a length scale of $\approx$3~au, which is just under the achieved synthesised beam size. This resolution is sufficient to detect particularly large or deep structures, however.

Sz~65 shows a low-contrast ring-gap structure that was confirmed through a visibility analysis. Inside of the ring, the brightness decrease is visible, but not particularly strong. We find a shallow gap, and the intensity structure we detect is symmetrical. The shallow nature of the gap might indicate that the structure is very narrow and has not been fully resolved by our observations. There are also physical explanations for a shallow gap; it might arise because the outer ring forms as a result of a leaky dust trap \citep{Zhu2012DUSTDISKS,Bosman2019ProbingEmission,Sturm2022TracingDisks}, where a weak pressure bump, or one that is ineffective at trapping millimetre-sized dust grains, has caused a pile-up of dust to create the outer ring in Sz~65, but is unable to completely prevent radial inward drift. If the structures are formed by planets, a low-contrast feature is likely. Discs with a low initial mass have a lower budget with which to form large planets, and the lower-mass planets are unable to carve gaps that are as deep as those that are carved by giant planets several times the mass of Jupiter, for example. In order to estimate the masses of planets in a protoplanetary disc using gap widths and contrasts, complex hydrodynamic models are required that involve numerous uncertain variables. We can calculate indicative values for upper limits on the maximum mass of planets in the system using scaling relations that arise from this modelling, however \citep{lodato2019,Ribas2023TheScales,Yang2023Multiple21}. For example, following \citet{lodato2019}, we can assume low viscosity ($\alpha < 0.01$) and that the width of the gap opened by a planet, $\Delta$, scales with the Hill radius of the planet, leading to a relation of the planet mass and the gap width, 
\begin{equation}
    M_{\rm planet} = \Big( \frac{\Delta}{k r} \Big)^3 \times 3 M_* ,
\end{equation}
where $\Delta$ is measured between the minimum brightness point in the gap and the maximum brightness of the ring exterior to it. k is a constant arising from the modelling, and it varies between 4.5-8 \citep{lodato2019,Rosotti2016}, M$_*$ is the stellar mass, and r is the radial position of the planet. For Sz~65, M$_*$=0.76 \citep{Alcala2017X-shooterObjects}, and we can measure the gap width as described by using the \textsc{frank} fit to the radial profile, which has a minimum gap brightness found at 20~au and a peak ring brightness at 24~au. When we account for variation in k, this means that an upper limit on the mass of a planet found is 0.04-0.2 M$_{\rm Jup}$, for the brightness variation we observe.  

Sz~66 is resolved by the new observations, but no new substructure is revealed. The subtraction of the \textsc{frank} model from the \emph{clean} image shows some residuals that might hint at asymmetry in the disc, but these strong residuals are very localised; they are located within one beam width. The peak residuals reach 7$\sigma$. These hints should be followed up with further high-resolution imaging at different wavelengths to allow a characterisation of the dust. Imaging at shorter wavelengths may be capable of revealing the morphology of the inner disc, as was achieved for discs around Herbig stars with near-IR imaging using PIONIER data by \citep{Kluska2020AUnits}.

For the molecular gas disc, the moment 0 maps of CO emission presented in Figure \ref{fig:gas_ims} reach relatively low peaks in signal-to-noise ratio, both of about 5$\sigma$. Whilst useful for constraining the disc size, deeper observations of these lines at a similar angular resolution are required to reliably confirm structures in the gas disc. In particular, maintaining a sufficient signal-to-noise ratio whilst increasing the spectral resolution of molecular line observations will aid in the kinematic analysis that explores the molecular gas distribution in greater detail.

\subsection{DSHARP sample comparison}
\label{sec:DSHARP_comp}

\begin{figure}[h]
    \centering
    \includegraphics[width=\linewidth]{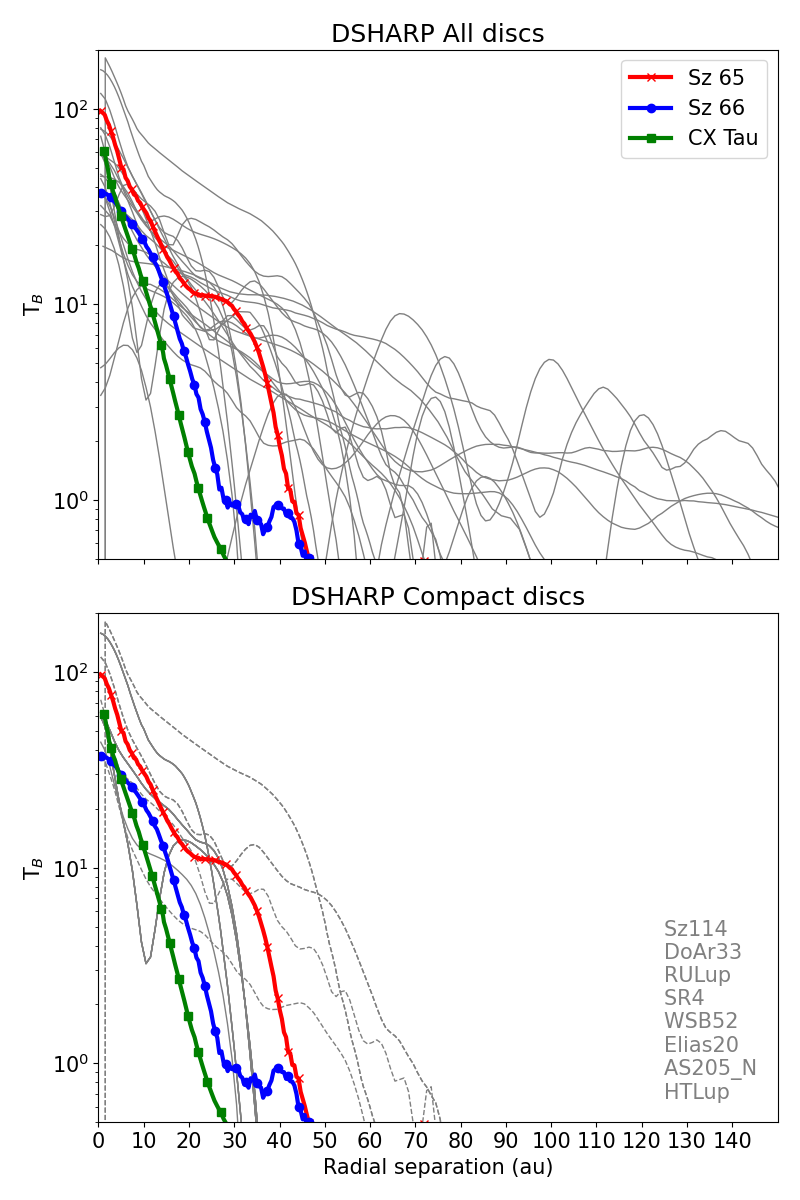}
    \caption{Brightness temperature as a function of radius calculated from the 1.3mm continuum images of Sz~65 (red) and Sz~66 (blue) compared with the results of the DSHARP survey (grey). In addition, we plot the radial profile of the well-resolved compact disc CX Tau \citep{Facchini2017}. All DSHARP discs are plotted in the top panel. Only discs classified as compact are plotted in the lower panel and are listed on the right side. Discs with a derived disc size $>$ 50~au are plotted with dashed lines.}
    \label{fig:comp_DSHARP}
\end{figure}

To assess to which extent our compact discs are comparable to massive and structured discs, we compared the radial profile of Sz~65 and Sz~66 to the results of the DSHARP survey, which contains discs with a variety of substructure.
The brightness temperature, T$_{\rm B}$, of the continuum radial profile was calculated and plotted alongside those from the DSHARP survey achieved with a similar resolution (DSHARP, $\approx$5~au, vs. this work, $\approx$3~au) in Figure \ref{fig:comp_DSHARP}, as was done in \citet{Facchini2019HighDisk} for the disc CX Tau. 

The top panel in Figure \ref{fig:comp_DSHARP} compares the T$_{\rm B}$(R) profile for Sz 65 and Sz 66 with the DSHARP sample and CX Tau \citep{Facchini2019HighDisk}. The difference in the radial extent of the compact discs is very clear.
The highly structured systems show deep rings and gaps and a much greater radial extent. In these outer regions, the pressure scale height is larger, and stable substructures are therefore broader radially. In the inner disc, these features are less frequent, with some exceptions on which we focus below. In the top panel of Figure \ref{fig:comp_DSHARP}, Sz~65 and Sz~66 have a lower brightness temperature than most DSHARP discs. This may be due to a decreased column density, a decreased temperature, or fewer millimetre-sized grains. 

A more like-for-like comparison can be made between the compact discs and the compact DSHARP discs alone.
In the bottom panel of Figure \ref{fig:comp_DSHARP}, we plot the T$_{\rm B}$ profile of Sz~65 and Sz~66 against the most radially compact discs from the DSHARP survey. In this grouping, Sz~65 no longer appears as an outlier, with a similar radial extent and T$_{\rm B}$ as the other systems. Sz~66, however, is more similar to CX Tau, and they are by far the two most radially compact objects. 
This group of objects is one of the best-resolved compact discs observed by ALMA so far. The substructure features we see are faint contrast gaps that could be caused by a leaky dust trap that cannot efficiently retain millimetre-particles in the outer disc.

\subsection{Structure in compact discs}

We present high-resolution observations of Sz~65 and Sz~66 for the first time. The structure we detect in Sz~65 supports the claim that substructure is ubiquituous at sufficiently high angular resolution. Sz~66, however, remains compact and lacks clearly identifiable substructures even at high resolution.
Our observations of two compact discs and the comparison with the most compact members of the DSHARP sample do not support the interpretation that compact discs in large surveys are simply scaled-down versions of the extended highly structured cases, that is, with smaller rings at shorter radii. This conclusion is shared by a study of a faint compact disc CX Tau by \citet{Facchini2019HighDisk}, which are resolved at 1.3mm with ALMA down to 5~au. 
Figure \ref{fig:comp_DSHARP} suggests that Sz~65 and Sz~66 have more in common with the DSHARP compact discs (in terms of radial extent) than they do with the wider DSHARP sample. 
We might then speculate that the former falls into a category of discs that lack efficient dust traps in the outer disc, whilst the latter belong to a category in which the millimetre grains have been retained in the outer disc. CX Tau was found to have a high gas-to-dust size ratio, which the authors interpreted as indicating efficient radial drift, which would support this analysis. \citet{Pinilla2021A1} analysed an example that might represent the other category, in which a planet-induced dust trap can explain the deep and wide cavity ($\approx$20~au) around the low-mass star CIDA 1.
We must also consider the possibility that compact discs simply formed very small, and so there is no need to invoke further processes during the evolution of the disc. Large gas discs such as we find in Sz~65 disagree with this scenario, but for Sz~66, this is perhaps a more plausible origin. Extremely compact discs have been identified in Lupus by \citep{vanderMarel2022High-resolutionLupus}, who reported that Sz~104 and Sz~112 have radii of 2.5 and 3.5~au, respectively. 

Categorising discs into the two groups of compact discs and DSHARP-like structured objects is convenient here to assess discs with similar properties, but it does not necessarily help us to understand the physics of the individual systems involved or their planet-forming ability. 

We can examine the link between compact discs and planet formation by considering that late-type stars, and M dwarfs such as Sz~65 in particular, are associated with a higher occurrence of transiting planets than Sun-like stars. They also show discs with substructure. \citet{Kurtovic2021SizeRegion} identified structures in the dust discs around three of the six very low mass stars in Taurus with 0$\farcs$1 resolution with ALMA at Band 7. Modelling of the planet formation around M dwarfs by \citet{Mulders2021WhyPlanets} is consistent with observed statistics, where the authors find the formation of hot super-Earths to be most likely around low-mass stars, whilst higher-mass stars are more likely to form a gas giant, and that the formation of this first giant can then interrupt the pebble flow into inner regions, inhibiting further formation. These results provide a physical mechanism that could explain some of the suggestions above concerning a drift-based evolution in comparison to one that is dominated by the dust traps, and as is discussed in \citet{vanderMarel2021ADemographics}. 
In a disc that has not created a giant planet early, we should also take note of the correlation between stellar mass and disc mass \citep{Andrews2013THEHOSTS,Ansdell2016, Pascucci2016ARELATION}. Discs around less massive stars are less massive, and so do not have the potential to birth multiple massive planets that can carve multiple gaps. The dust mass estimate made here is 16.1~M$_{\oplus}$, whilst for HD~163296, using the same method as above and the integrated flux given in \citet{Andrews2018}, we calculate that the dust mass could be as high as $\approx$187~M$_{\oplus}$. These masses calculated in the same manner indicate that the disc around the $\approx$2 M$_\odot$ star contains significantly more millimetre dust, seemingly by order(s) of magnitude. Less massive discs will lead to less massive planets, or indeed fewer planets. Less massive planets will naturally result in less clearly defined features \citep{Rosotti2016}, and often require low turbulence in the disc to form structures such as rings at all \citep{Kuwahara2022DustDisks}. Analysing observations of a total of 700 discs,  \citet{vanderMarel2021ADemographics} found that $\approx 10\%$ of the discs are structured for the given resolution of the archival observations, and that the proportion of discs in which rings or cavities can be identified increases in discs with a greater dust mass. It is therefore perhaps expected that we see deep gaps in HD~163296 at millimetre wavelengths \citep{Isella2018TheDisk}, for instance, but in this work, we detect only a low-contrast gap in Sz~65.

Fundamentally, the contrast in annular features, that is, the depth of a gap or the sharpness of a ring, is a function of the efficiency of the initial dust trap in being able to trap particles with a size relevant to the observing wavelength, and for how long the trap can survive. Optical depth effects can also lead to a variation in emission brightness and may also be at work in these systems. Observations at longer wavelength will be a key tool for investigating this possibility and can be achieved in the future using ALMA with receivers in Band 1 or 3 (2-6 - 8.6 mm) or the future ngVLA \citep[][3~mm-1~cm]{Ricci2018ScienceNgVLA}.
Strong gas pressure bumps lead to spectacular annular structures, but short-lived or weak bumps may instead lead to cavities with much dust within, as we find here, or very radially compact discs. Equally, turbulent discs with high $\alpha$ viscosity will work to limit these structures, whilst low-turbulence discs allow them to survive longer.
In a single system, when and where in a disc a pressure bump forms will dictate how the dust evolution in the disc proceeds from that point. Many suggested mechanisms for forming such bumps are available in the literature, each of which works over different length scales and at different points in the lifetime of a disc. This provides a plethora of potential evolutionary pathways that often depend on local system-specific properties. For example, Sz~65, Sz~66, and SR4 (included here as part of the DSHARP sample) are all considered compact discs, one of which has a low-contrast substructure, one has no obvious structure at all, and one has a very clear inner gap. Only for much greater sample sizes of spatially well-resolved compact discs or by considering alternative unifying properties other than disc size and frequency of structure might an overall trend emerge. Promising avenues for future work might be to consider the extent of the chemical evolution expected in compact discs \citep{Miotello2021CompactDisks}, a multi-wavelength approach to compact discs, or a focus on achieving spatially and spectrally well-resolved observations of molecular gas as well as dust continuum emission. 

\section{Conclusions}

We have presented high-resolution ALMA observations of two compact discs in Lupus, Sz~65 and Sz~66. 
We carried out an in-depth analysis of the dust continuum emission in Sz~65 in order to detect signs of substructure in the disc. Through non-parametric visibility modelling, we found evidence of a shallow gap at about 20~au, with a ring-like increase in emission at the edge of the disc. This gives proof in this individual case of structure in a compact disc, but not a structure that is so readily apparent as those found in the DSHARP survey, for example. The ring and gap in Sz~65 that we find here may have arisen due to a leaky dust trap that cannot efficiently retain millimetre-sized dust in the outer disc. Non-parametric visibility fitting finds a possible feature at 6~au for which further observational confirmation is necessary.
Sz~66 is resolved spatially, but is found to be very radially compact, with some evidence of an asymmetric dust distribution seen when a disc model fitted to the visibilities is subtracted from the image, but an otherwise smooth or structureless profile, much like unresolved compact discs in previous surveys. The observations do not support the hypothesis that these two discs are simply radially smaller size-scaled versions of the bright radially extended DSHARP discs that show multiple substructures. Sz~65 and Sz~66 instead show a clear outer edge of millimetre~material, and only Sz~65 shows evidence of a ring, but no evidence of millimetre structures on wider scales. 

Molecular gas in the discs is detected reliably only after the application of a uv taper of 0$\farcs$07 to our observations. With these data, we constrain the size of the molecular gas disc with an improved spatial accuracy in comparison to previous observations. Both Sz~65 and Sz~66 have gas discs that are more extended than their dust discs. The ratio of R$_{\rm gas}/$R$_{\rm dust}$ for Sz~65 is 5.3$\pm$0.9, suggesting a disc that has undergone dust evolution. For Sz~66, the ratio is 2.9$\pm$1.0, but the relative uncertainties are high because the disc is less well resolved spatially. Nevertheless, even within the uncertainties, the current data suggest that this disc has a ratio R$_{\rm gas}/$R$_{\rm dust}$ lower than the indicative value of 4, perhaps suggestive of truncation, or of an initial disc that was also very compact. Tests of the statistical likelihood of disc truncation due to binary interaction suggest that truncation can only have been a significant influence if the eccentricity of the Sz~65 - Sz~66 orbit is very high. Future observations are needed to constrain these orbital properties further.

The categorisation of discs into structured and unstructured strongly relies on the observing set-up used to take the data. Structure may always exist on smaller scales or may be visible at different wavelengths that are not apparent in a particular image. It is more physically useful to think about drift-dominated and trapping-dominated evolution of the dust. The ability of a disc to form and sustain a gas pressure bump is the main driver of annular structures such as rings and gaps. Asymmetric structure such as spiral arms or clumps often depend on the dynamics between star and planet, a star and a binary partner, or fly-by interaction. These processes may still occur on smaller scales because a typical definition of any such features is yet to be found. The comparison of Sz~65 and Sz~66 with the DSHARP radial profiles showed that compactness does not necessarily lead to a lack of structure. There does appear to be a link between radial size of the disc and the clarity of these structures, for example, deeper cavities in the extended DSHARP discs. We urge continued research into compact discs, however, in order to further investigate unifying properties amongst this population, for example, investigating predictions of rapid chemical evolution in compact discs \citep{Miotello2021CompactDisks}.

\bibliographystyle{aa}
\bibliography{references_update.bib} 

\begin{acknowledgements}
This paper makes use of the following ALMA data: ADS/JAO.ALMA\# 2017.1.00569.S, 2018.1.00271.S, 2018.1.01458.S, 2019.1.01135.S, 2015.1.00222.S. ALMA is a partnership of ESO (representing its member states), NSF (USA) and NINS (Japan), together with NRC (Canada), MOST and ASIAA (Taiwan), and KASI (Republic of Korea), in cooperation with the Republic of Chile. The Joint ALMA Observatory is operated by ESO, AUI/NRAO and NAOJ. This research has made use of the SIMBAD database, operated at CDS, Strasbourg, France. TJH is funded by a Royal Society Dorothy Hodgkin Fellowship. M.B. received funding from the European Research Council (ERC) under the European Union's Horizon 2020 research and innovation programme (PROTOPLANETS, grant agreement No.~101002188). This work made use of Astropy:\footnote{http://www.astropy.org} a community-developed core Python package and an ecosystem of tools and resources for astronomy \citep{astropy:2013, astropy:2018, astropy:2022}. The authors would additionally like to thank Megan Ansdell for making available her fits files and scripts pertaining to archival data on Sz~65. Thanks also to participants of the "Planet and binary formation in gravitationally unstable protoplanetary discs in the high-resolution era" conference for productive and inspiring discussions.
\end{acknowledgements}

\appendix
\renewcommand{\thefigure}{A\arabic{figure}}
\setcounter{figure}{0}
\section*{Appendix}

\begin{figure*}[h]
    \centering
    \includegraphics[width=0.9\textwidth]{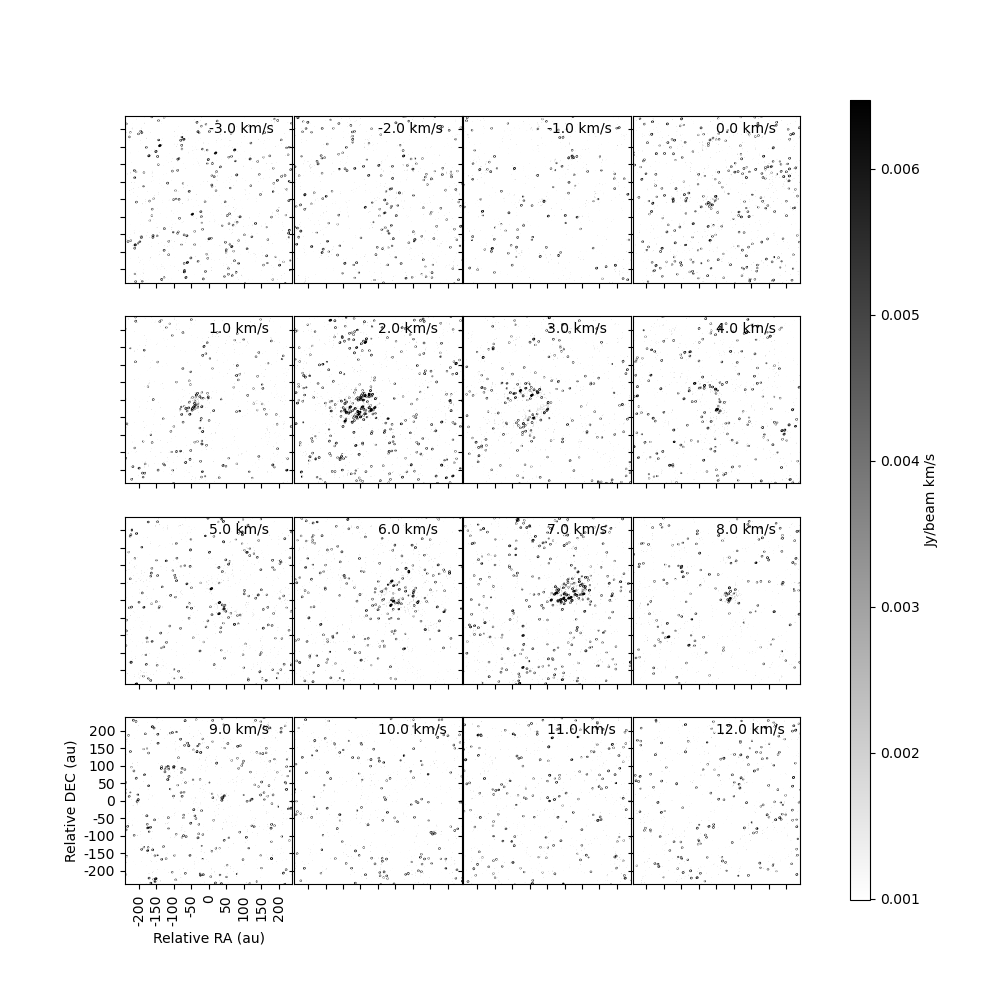}
    \caption{Channel maps of $^{12}$CO(2-1) in Sz~65 at the native resolution with a synthesising beam 0$\farcs 026 \times 0\farcs 017$. The contours are 3,5,7$\times \sigma$, where $\sigma$=0.89 mJy/beam km/s.  }
    \label{fig:chans_notaper}
\end{figure*}

\begin{figure*}[h]
    \centering
    \includegraphics[width=0.9\textwidth]{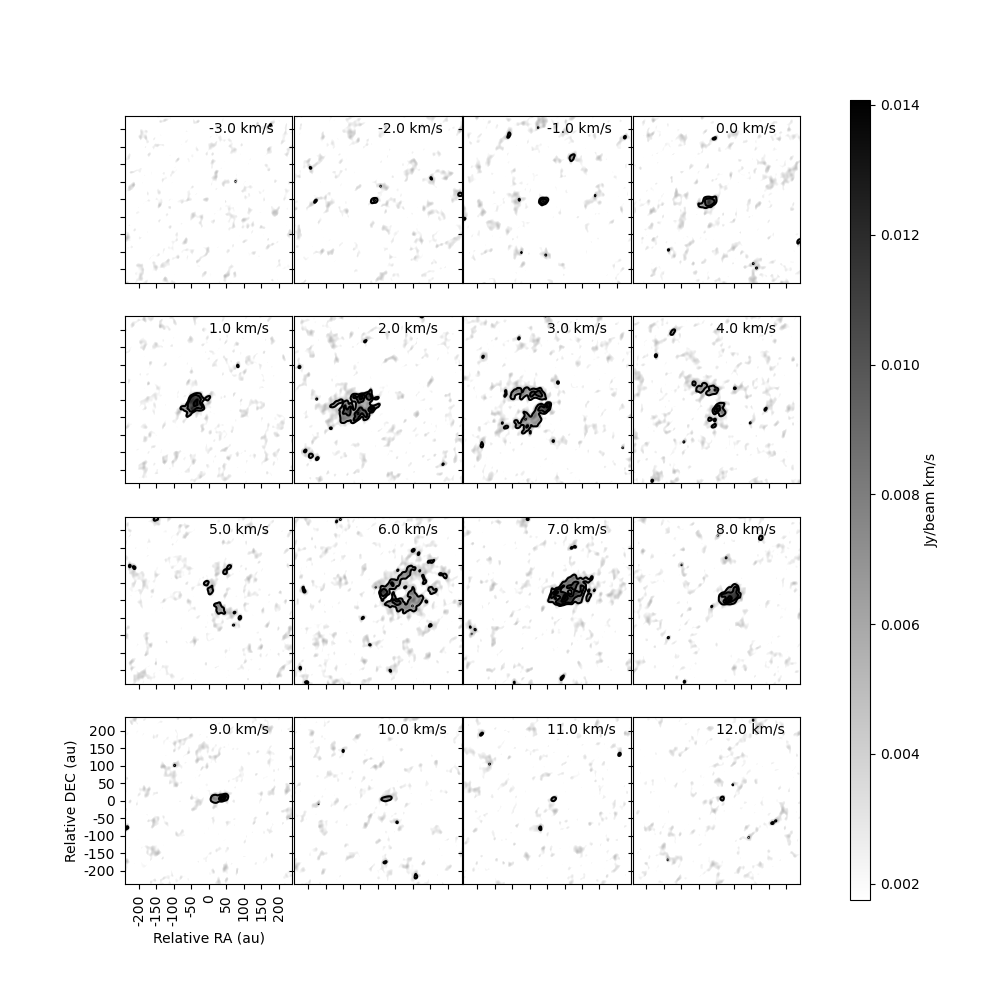}
    \caption{Channel maps of $^{12}$CO(2-1) in Sz~65 after the application of 0$\farcs$1 taper, resulting in a synthesising beam 0$\farcs 101 \times 0\farcs 079$. The contours are 3,5,7$\times \sigma$, where $\sigma$=1.9 mJy/beam km/s.  }
    \label{fig:chans_taper}
\end{figure*}

\begin{figure*}
    \centering
    \includegraphics[width=0.9\textwidth]{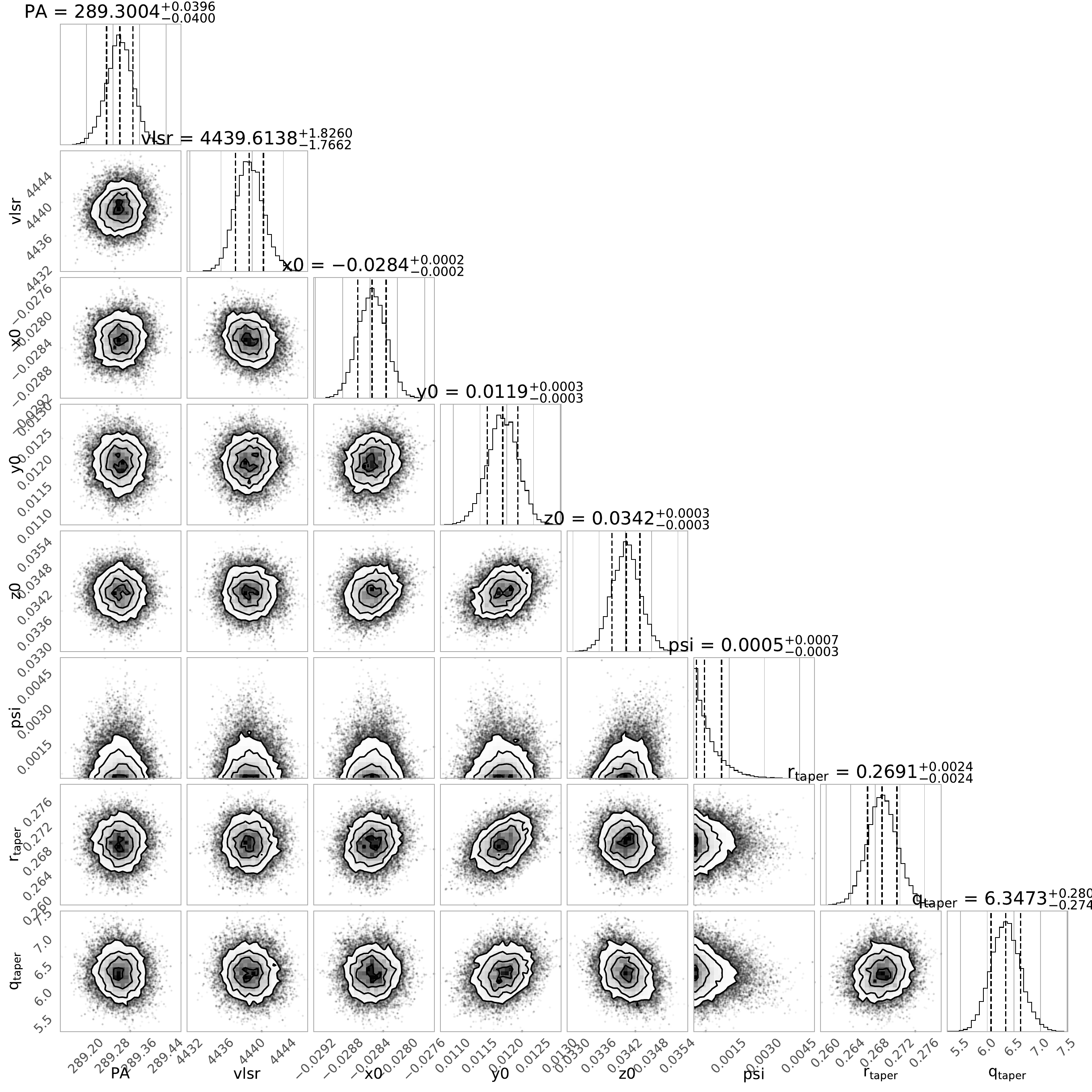}
    \caption{Corner plot showing the MCMC results for the rotational model fit to the $^{12}$CO observations, achieved by fixing the inclination to the continuum value of 63.76$^{\circ}$, using a stellar mass of 0.76~M$_{\odot}$ \citep{Alcala2017X-shooterObjects} and a distance derived from Gaia EDR3 153.1 pc.}
    \label{fig:corner65}
\end{figure*}

\begin{figure*}
    \centering
    \includegraphics[width=0.9\textwidth]{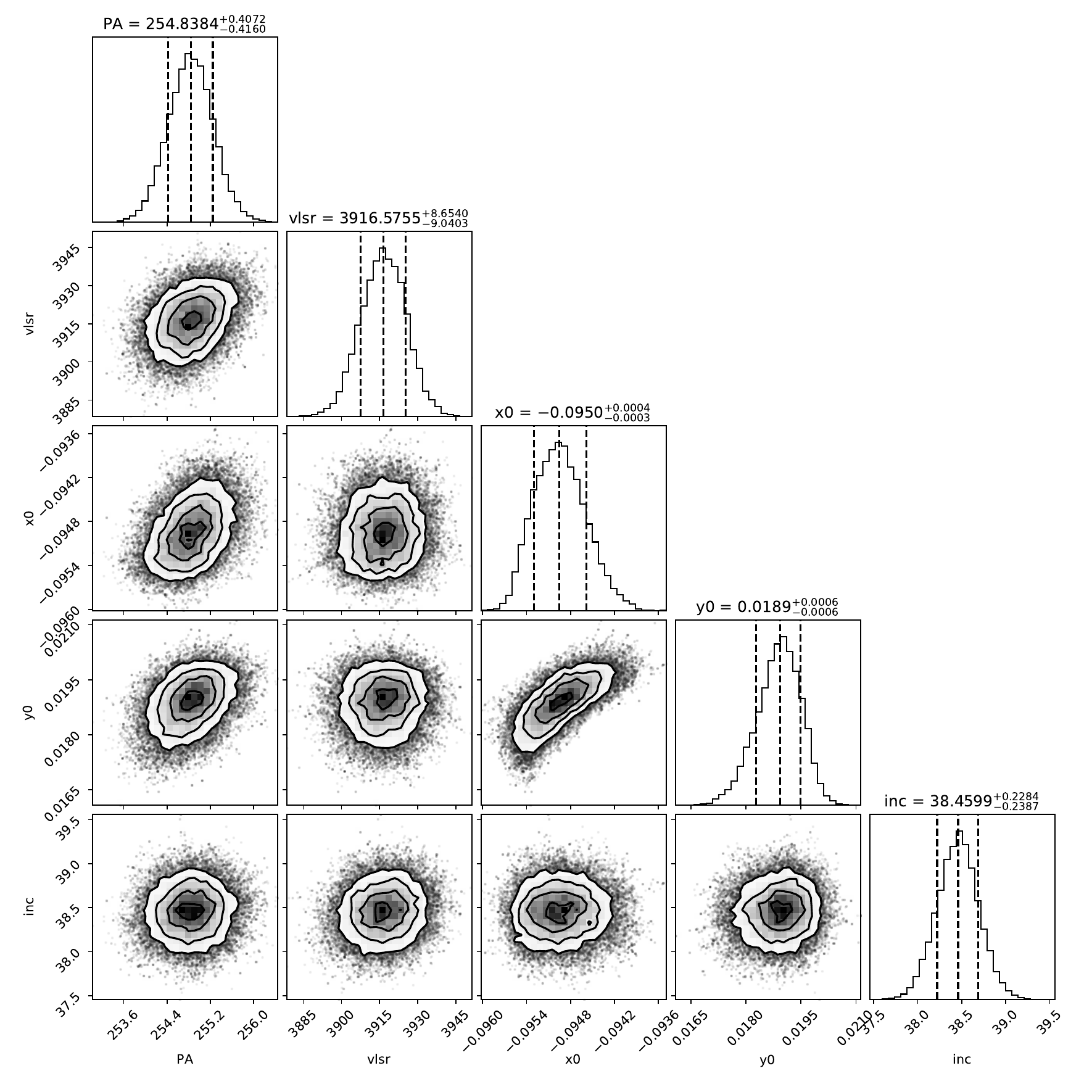}
    \caption{Corner plot showing the MCMC results for the rotational model fit to the $^{12}$CO observations, achieved by fixing the stellar mass of 0.31~M$_{\odot}$ \citep{Alcala2014X-shooterObjects} and a distance derived from Gaia EDR3 156.0 pc.}
    \label{fig:corner66}
\end{figure*}

%
%

\end{document}